\documentclass[reprint,aip,jcp,a4paper,floatfix]{revtex4-1}

\usepackage{amssymb,amsmath}
\usepackage{graphicx}

\begin{document}

\title{Functionality in single-molecule devices: Model
  calculations and applications of the inelastic electron tunneling
  signal in molecular junctions}

\author{L.~K. Dash}
\email{louise.dash@york.ac.uk}
\affiliation{Department of Physics, University of York, Heslington, York YO10 5DD, UK}
\affiliation{European Theoretical Spectroscopy Facility (ETSF)}

\author{H.~Ness}\address{Department of Physics, University of York, Heslington, York YO10 5DD, UK}
\affiliation{European Theoretical Spectroscopy Facility (ETSF)}

\author{M.~J. Verstraete}
\affiliation{Institut de Physique, Universit\'e de Li\`ege, B-4000 Sart Tilman, Belgium}
\affiliation{European Theoretical Spectroscopy Facility (ETSF)}

\author{R.~W. Godby}
\affiliation{Department of Physics, University of York, Heslington, York YO10 5DD, UK}
\affiliation{European Theoretical Spectroscopy Facility (ETSF)}

\begin{abstract}
  We analyze how functionality could be obtained within
  single-molecule devices by using a combination of non-equilibrium
  Green's functions and \emph{ab-initio} calculations to study the
  inelastic transport properties of single-molecule junctions.  First
  we apply a full non-equilibrium Green's function technique to a
  model system with electron-vibration coupling. We show that
  the features in the inelastic electron tunneling spectra (IETS) of
  the molecular junctions are virtually independent of the nature of
  the molecule-lead contacts.  Since the contacts are not easily
  reproducible from one device to another, this is a very useful property.
  The IETS signal is much more robust versus modifications at the
  contacts and hence can be used to build functional nanodevices.
  Second, we consider a realistic model of a organic conjugated molecule. 
  We use \emph{ab initio} calculations to study how the
  vibronic properties of the molecule can be controlled by an external
  electric field which acts as a gate voltage.  The control,
  through the gate voltage, of the vibron frequencies and (more
  importantly) of the electron-vibron coupling enables the construction of 
  functionality: non-linear amplification and/or switching is obtained
  from the IETS signal within a single-molecule device.
\end{abstract}

\maketitle


\section{Introduction}
\label{sec:introduction}

Single-molecule electronics has shown significant progress in recent
years.  Aviram and Ratner first proposed in the 1970s electronic
devices where a single molecule functions as the active element
\cite{aviram:1974}. Since then a variety of interesting effects has
been observed in the transport properties of single molecules,
including rectification, negative differential resistance, and
switching.  The molecules usually employed are organic conjugated
molecules with strongly delocalized $\pi$-electrons along the
molecular backbones (and eventually functional chemical groups acting
as electron donors or acceptors), and are natural candidates for
bottom-up assembly of electronic devices.  Chemical synthesis is a
massively parallel method to create identical molecules by the
mole. Their structures and electronic properties can be tailored
virtually at will and can be characterized by several spectroscopic
and structural probe techniques. Thus they are ideal building blocks
for electronic devices.

However, to make even the simplest molecular electronic device---a
single molecule connected between two electrodes---presents several
practical difficulties.  The fabrication of the nanojunction is a
challenge because of the difference in scales between the molecule,
the electrodes, and the small gap between them. Moreover, it is even
more difficult to verify that the intended molecule is indeed in the
junction, that it is well connected, and that it is oriented in the
expected manner with respect to the electrodes.  Several methods are
used to create such nanojunctions (e.g. mechanical break junctions
\cite{Reed:1997,Reichert:2003}, crossed wires \cite{Kushmerick:2004},
and scanning probe microscopy \cite{Cui:2001,Venkataraman:2006}).
Each technique has its own advantages. For example, some of them can
be cooled to low temperature, others provide a good control of the
electrode spacing, or provide images of the contact area.  However,
the most crucial and as yet uncontrolled point is the nature of the
contacts between the molecule and the electrodes.  This is pivotal,
since the transport properties of nanojunctions do not depend just on
the intrinsic properties of the molecules, but also on the nature of
the contacts.  Although there is an increasing amount of research into
the control of these contacts by tailoring the end groups of the
molecule anchoring on the surface
\cite{Venkataraman:2006,Baheti:2008,DellAngela:2010}, there is still a
great deal of uncertainty about the specific nature of the contacts.
These change from device to device, or even fluctuate during a single
experiment, because of thermal effects (atom migration at the
electrode surface) or of current-induced effects. Hence it is
extremely difficult to produce molecule junctions with reproducible
properties.

Other properties of current-carrying molecular junctions are also
important to understand, such as the nature and the effects of
electron-vibron coupling on the molecule. Such a coupling can
modify the transport properties of the junctions at a particular
threshold voltage and is also responsible for heating effects in the
junctions.  Several groups have already performed inelastic electron
tunneling spectroscopy (IETS) on molecular junctions to obtain a
signature from the vibration modes of the molecules in the
nanojunction \cite{Hipps:1993,Liu:2004,Kushmerick:2004,
  Yu:2004,Wang:2004,Yu:2006,Chae:2006,Troisi:2006c,Beebe_JM:2007,Okabayashi:2008,Kim:2010}.

The realization of a true single-molecule device, with three
terminals, has been achieved recently \cite{Song:2009}. The transport
properties through the source and drain are directly modulated by an
external gate voltage as in a conventional field-effect transistor.
However, in the molecular transistor, the gate voltage modifies the
molecular orbital energies. In Ref.~\onlinecite{Song:2009}, the
authors also studied the effects of the gate voltage and temperature
on the IETS signal.  They found that depending on the nature of the
molecule, the IETS signal is nearly independent of the gate voltage
(for a molecule with a $\sigma$-saturated alkyl backbone), however
significant modifications of the IETS features' intensity and
lineshape exist for $\pi$-conjugated molecules with an aromatic ring.
Their results indicate that both conductance and IETS signal can be
modified and controlled by a gate voltage.

Full \emph{ab-initio} calculations for a realistic single-molecule
with three terminals (source, drain and gate) are extremely
computationally demanding, especially for self-consistent calculations
and/or calculations for realistic non-equilibrium conditions.
Non-equilibrium transport properties have been calculated for
realistic systems of single-molecules connected to two terminals by
using density-functional theory
\cite{Hirose1994,DiVentra2000,Taylor2001,Nardelli2001,Brandbyge:2002,Gutierrez2002,
  Frauenheim:2002,Xue2003,Louis+Palacios:2003,Thygesen2003,Garcia-Suarez2005}.
Calculations in the presence of a third gate electrode are rare
\cite{Ghosh2004, Zahid:2005,Vasudevan:2010} and usually use model
Hamiltonians (i.e.~tight-binding or extended H\"uckel).  In this paper
we address the problem of functionality in three-terminal
single-molecule devices by using a combined (two step) theoretical
framework. We discuss the use of both the electronic and the vibronic
properties of the molecule to achieve a device with functionality.

The first step consists of a detailed analysis of the full
non-equilibrium transport properties of a model system. We study the
effects of the contacts between the molecule and the source and the
drain on the conductance and the IETS signal. The presence of the gate
voltage is incorporated in the position of the molecular levels and
generates transitions between different transport regimes (resonant
and off-resonant).  We use a full non-equilibrium Green's function
(NEGF) technique \cite{Dash2010b,Dash2011,Ness2010, Mii:2003,
  Frederiksen:2004b, Galperin:2004b, Pecchia:2004b, Ryndyk:2006,
  Sergueev:2005, Viljas:2005, Yamamoto:2005, Cresti:2006, Kula:2006,
  Vega:2006, Ryndyk:2006, Troisi:2006b, Vega:2006, Galperin:2007,
  Ryndyk:2007, Schmidt:2007, Troisi:2007, Asai:2008, Benesch:2008,
  Paulsson:2008, Egger:2008, McEniry:2008}.  In order to explore a
wide range of experimental contact geometries, we first study a model
system, in which two parameters (the contact strength and the
corresponding fractions of potential drop)
\cite{Galperin:2004b,Datta:1997} characterize the molecule-lead
coupling.  We show that the IETS signal is much less sensitive to the
nature of the contact than the conductance itself.

The second step consists of applying the principle found in the first
step to a more realistic model. We study the influence of an external
electric field on the electronic and vibronic properties of a
realistic molecular system based an ethynylphenyl-based backbone by
using \emph{ab-initio} calculations \cite{abinit:2009}.  Having shown,
for the model system, that the IETS signal is nearly independent of
the nature of contacts, we concentrate on an isolated molecule and
extract the relevant physical quantities (vibration frequencies and
electron-vibron coupling matrix elements) in the presence of the
electric field which simulates the presence of a gate electrode.

Finally we discuss and show the possibility of using the IETS signal,
rather than the current or the conductance, to form functional
nanodevices with switching, non-linear amplification or sensor
functionality by using our \emph{ab-initio} calculations for the
conjugated molecule.

\section{Non-equilibrium transport and the effect of the contacts}
\label{sec:model}

We use a NEGF technique to calculate self-consistently the full
non-equilibrium inelastic properties of a molecular junction.  Using a
model system to reduce these calculations to a tractable size, we
concentrate on a single molecular level coupled to a single
vibrational mode.  A full description of our methodology is provided
in Refs.~[\onlinecite{Dash2010b,Ness2010,Dash2011}].

This approach allows us to study how variations at the contacts modify
the transport properties of the junction.  These include
experimentally uncontrollable modifications of the molecule-lead
coupling due to the experimental environment---such as thermal
fluctuation, diffusion of atoms at the surface of the leads, variation
of the gap of the nanojunction, the presence of impurity molecules
around the nanojunction, etc.  The consequence of these modifications,
and the reason they limit the reproducibility of functionalized
junctions, is to change the geometry of the contacts, and therefore
the strength of electronic coupling between the molecule and the
leads, and the corresponding potential drops at the contacts.

In our model, the central region of the nanojunction, i.e.\ the molecule, is
described by a simple electron-vibron coupling Hamiltonian
\begin{equation}
  \label{eq:H_central}
  H_C = \varepsilon_0 d^\dagger d + \hbar \omega_0 a^\dagger a +
  \gamma_0 (a^\dagger + a) d^\dagger d,
\end{equation}
where one electronic molecular level $\varepsilon_0$ and one vibration
mode with energy $\omega_0$ are coupled together via the coupling
constant $\gamma_0$. The electron and phonon creation/annihilation
operators are $d^\dagger$/$d$ and $a^\dagger$/$a$ respectively.

The central region is then coupled to two non-interacting Fermi seas
each at their own equilibrium, characterized by the left and right
Fermi levels $\mu_L$ and $\mu_R$, via two hopping matrix elements
$t_{0L}$ and $t_{0R}$ which represent the strength of electronic
coupling between the molecule and the leads.

All the properties of the nanojunctions (electronic density, spectral
functions, current density) are determined from the knowledge of the
NEGF of the central region \cite{Dash2010b,Ness2010,Dash2011}. For
example, the retarded Green's function of the central region is given by
\begin{equation}
  \label{eq:GrDyson}
  G^{r}(\omega) = [ \omega -\varepsilon_0 - 
\Sigma_L^{r}(\omega) +
  \Sigma_R^{r}(\omega) + \Sigma_{e-{\rm vib}}^{r}(\omega) ]^{-1},
\end{equation}
where $\Sigma_\alpha^{r}$ is the self-energy arising from the $\alpha=L,R$
lead and $\Sigma_{e-{\rm vib}}^{r}$ is the self-energy arising form the 
electron-vibron interaction. The latter is calculated from Feynman diagram
expansion of the interaction \cite{Dash2010b}. In this section, we calculate
$\Sigma_{e-{\rm vib}}$ from the lowest order expansion of the electron-vibron 
interaction (first Born or Hartree-Fock level of approximation) \cite{Dash2010b,Dash2011}.
The calculations are performed self-consistently \cite{Dash2010b,Lu:2007}.

\begin{figure}
\begin{center}
\includegraphics[clip,width=\columnwidth]{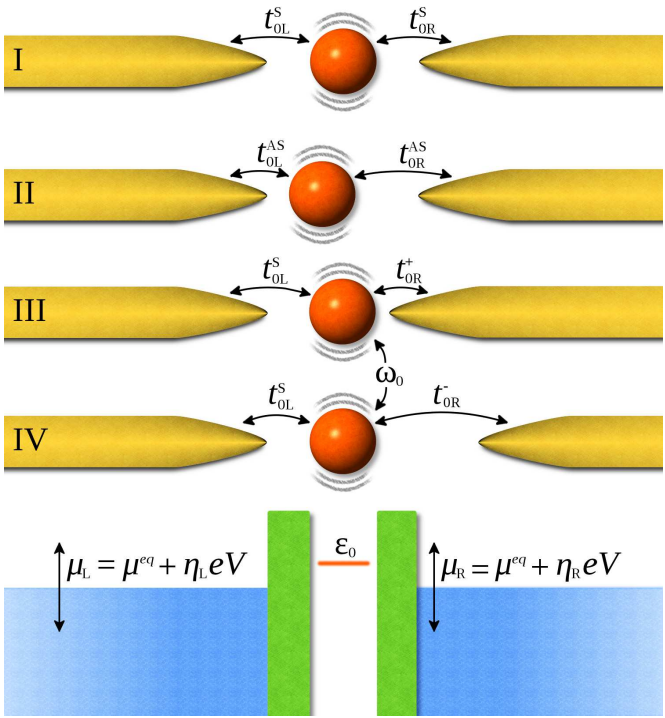}
\end{center}
\caption{Schematic representation of our model system. Case (I) shows
  a symmetric coupling at the contacts, with $t^S_{0L} =
  t^S_{0R}$. Cases (II) to (IV) are asymmetric
  junctions. In Case (II) the total gap is kept constant, but there is
  a variation in the position of the central region,
  i.e. $t^{AS}_{0L}+t^{AS}_{0R}=t^S_{0L}+t^S_{0R}$.  In Case (III) the
  total gap is compressed,
  $t^{AS}_{0L}+t^{AS}_{0R}<t^S_{0L}+t^+_{0R}$, while in Case (IV) it
  is expanded, $t^{AS}_{0L}+t^{AS}_{0R}>t^S_{0L}+t^-_{0R}$. }
\label{fig:OurSystem}
\end{figure}

Within our NEGF model, the electrostatics are not solved
self-consistently with the non-equilibrium electron charge density,
and we thus have another degree of freedom for the potential drops at
the contacts.  At equilibrium, the whole system has a single and
well-defined Fermi level $\mu^{\rm eq}$.  Out of equilibrium, a finite
bias is applied across the junction, giving Fermi levels
$\mu_{L,R}=\mu^{\rm eq}+\eta_{L,R} eV$.  Following
Ref.~[\onlinecite{Datta:1997}], the fraction of electrostatic potential
drop at the left contact is $\eta_L=+\eta_V$ and $\eta_R=-(1-\eta_V)$
at the right contact, with $\eta_L-\eta_R=eV$ and $\eta_V \in [0,1]$,
as shown in Fig.~\ref{fig:OurSystem}.

The parameter $\eta_V$
characterizing the potential drop at the contact is related to the
other parameters of the junction, as we can see from the following:
Let us consider three typical cases.  For the first, when the coupling
at the contacts is symmetric, i.e.\ when $t_{0L}=t_{0R}$, it is
reasonable to assume a symmetric potential drop, i.e.\ $\eta_V=1/2$.
In the second case, the coupling is very strong on one side, for
example $t_{0L} \gg t_{0R}$, the left Fermi level $\mu_L$ is pinned to
its equilibrium value $\mu^{\rm eq}$ and there is a large tunneling
barrier at the right contact where all the potential drop occurs,
i.e.\ $\mu_R=\mu^{\rm eq}+ eV$.  Finally, we have the opposite case,
when $t_{0L} \ll t_{0R}$, the right Fermi level $\mu_R$ is pinned to
$\mu^{\rm eq}$ all the potential drop occurs at the left contact,
i.e.\ $\mu_L=\mu^{\rm eq}+ eV$.

For intermediate cases, it is therefore reasonable to assume a
lowest-order linear dependence of the parameter $\eta_V$ with the
hopping integral parameters $t_{0L,R}$, and hence we choose the
following relation $\eta_V = t_{0R}/(t_{0L}+t_{0R})$.  We now
have a relation between the strength
of electronic coupling between the molecule and the leads and
the corresponding fractions of potential drop.
In reality the hoping integrals depend exponentially on the distance
between atoms and the dependence of the fractions of potential drop on
such a distance obeys a different relation. In principle, the later
depends on the dielectric properties of the fully connected molecular
junctions. Example of fractions of potential drop $\eta$ calculated
for realistic molecular junctions, with symmetric and asymmetric
coupling, are given in Ref~[\onlinecite{Chen:2010}].
 
In the following, we will
consider four different cases, as depicted in Fig.~\ref{fig:OurSystem},
corresponding to different possible modifications of a symmetric
nanojunction, and calculate their transport properties. However,
in all the four cases, the fractions of potential drop will be close
to the symmetric case.

We concentrate on the regime corresponding to tunneling through the
{\sc homo}-{\sc lumo} gap of an organic molecular junction, where the
molecular level is well above (or below) the equilibrium Fermi level
$\mu^{\rm eq}$). This is a typical behavior for a semiconducting-like
molecule sandwiched between two electrodes when the gate voltage is
small.

The calculations for the NEGF model system can be reduced to unitless
parameters, i.e.\ normalized by the hopping integral for the
semi-infinite one-dimensional leads.  We present the most relevant
results of our study for the following set of parameters:
$\epsilon_0=1.5,\omega_0=0.4,\gamma_0=0.35$ and two sets of coupling
to the leads: medium coupling with $t^S_{0L}=t^S_{0R}=0.27$ and weak
coupling with $t^S_{0L,R}=0.15$.  A wide range of parameters has been
explored, and this choice is the closest to those calculated from
first principles (see below).
 
The fluctuations in the nanojunction introduce variations of the
hopping matrix elements shown in Fig.~\ref{fig:OurSystem}: from a
symmetric coupling junction, we obtain an asymmetric junction when the
position of the molecule inside the gap is varied by a small amount
($10\%$). Hence the left and right hopping integrals are
$t^{AS}_{0L}=t^S_{0L}+10\%$ and $t^{AS}_{0R}=t^S_{0R}-10\%$
respectively---case (II).  We also consider two cases for which the
gap is modified according to $t^+_{0R}=t^{AS}_{0L}$ for a compression
of the gap---case (III)---and as $t^-_{0R}=t^{AS}_{0R}$ for a
expansion of the gap---case (IV).  The potential drop parameter
$\eta_V$ that determines the fractions of potential drop at each
contact is calculated from $\eta_V = t_{0R}/(t_{0L}+t_{0R})$ as we
have explained above. For our set of parameters, we get $\eta_V=0.50$
for case (I), $\eta_V=0.45$ for case (II), $\eta_V=0.5238$ for case
(III) and $\eta_V=0.4737$ for case (IV).

\begin{figure}[h]
\begin{center}
\includegraphics[clip,width=\columnwidth]{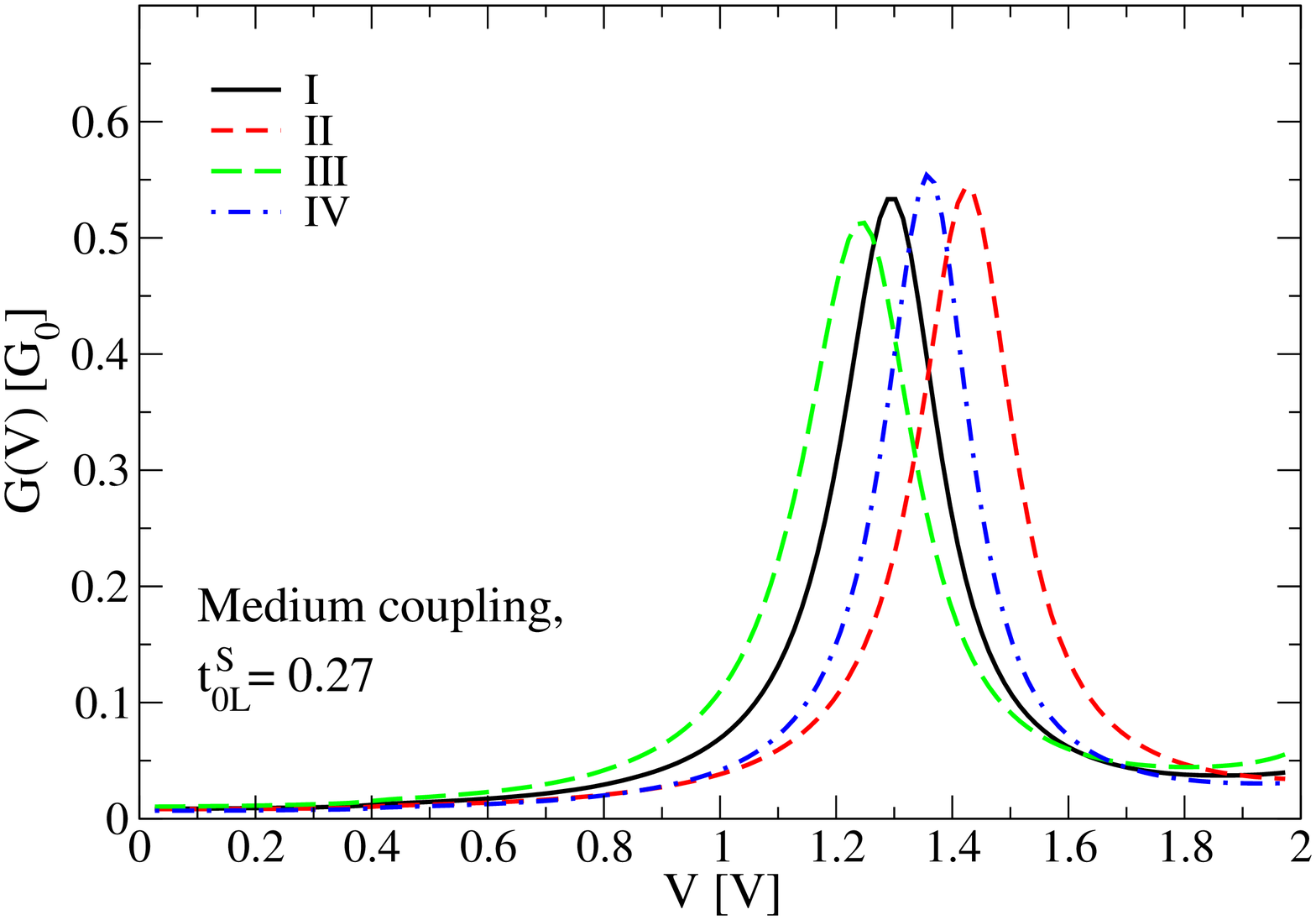} \\ 
\includegraphics[clip,width=\columnwidth]{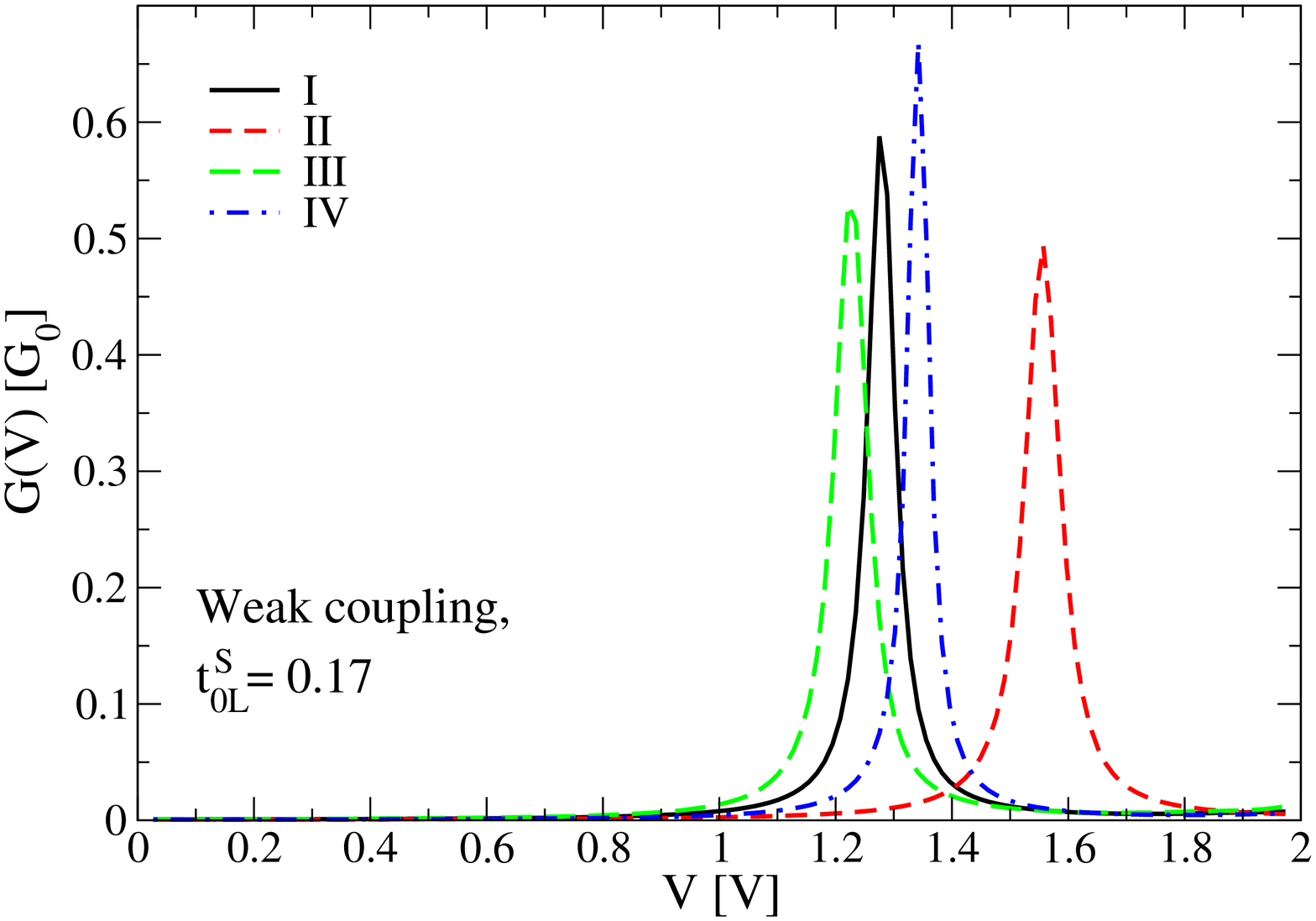}
\end{center}
\caption{Dynamical conductance $G(V)=dI/dV$ versus applied bias for the four cases
  depicted in fig.~\ref{fig:OurSystem}: symmetric (I) and asymmetric
  cases (II,III,IV). The top panel shows medium coupling to the leads
  ($t^S_{0L}=0.27$) and the bottom panel weak coupling
  ($t^S_{0L}=0.17$). Both sets of calculations show that, even for small
  variations of the potential drop parameter $\eta_V$, the position
  of the main conductance peak is strongly dependent on the fraction
  of potential drop at the contacts.}
\label{fig:conductance-curves}
\end{figure}

The corresponding conductance curves are shown in
Fig.~\ref{fig:conductance-curves}, and show one main conductance peak
for each configuration of the junction as expected.  The conductance
peak corresponds to a resonant transmission through the main
electronic level of the central region, renormalized by the
electron-vibron coupling, i.e.\ a peak at $\approx \tilde\epsilon_0
\sim \epsilon_0 - \gamma_0^2/\omega_0$.  It is clear from
Fig.~\ref{fig:conductance-curves} that the bias position of the peak
depends strongly on the value of the potential drop factor $\eta_V$
when all the other parameters of the central region are kept
unchanged. The width of the peak, proportional to $t^2_{0L}+t^2_{0R}$,
is not greatly affected by the $10\%$ fluctuation of the contacts for
the values of the parameters we have chosen.

\begin{figure}
\begin{center}
\includegraphics[clip,width=0.8\columnwidth]{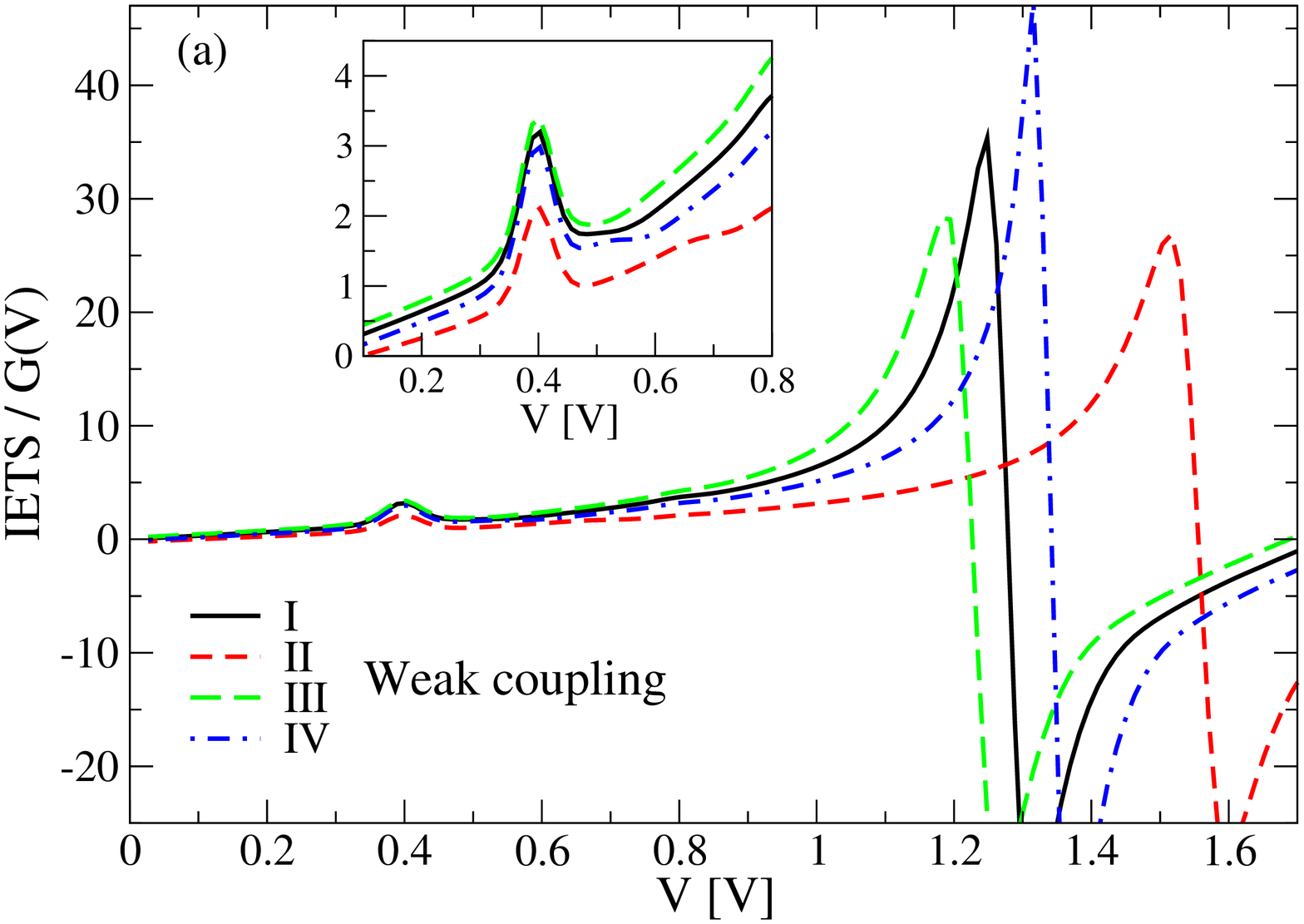} \\
\includegraphics[clip,width=0.8\columnwidth]{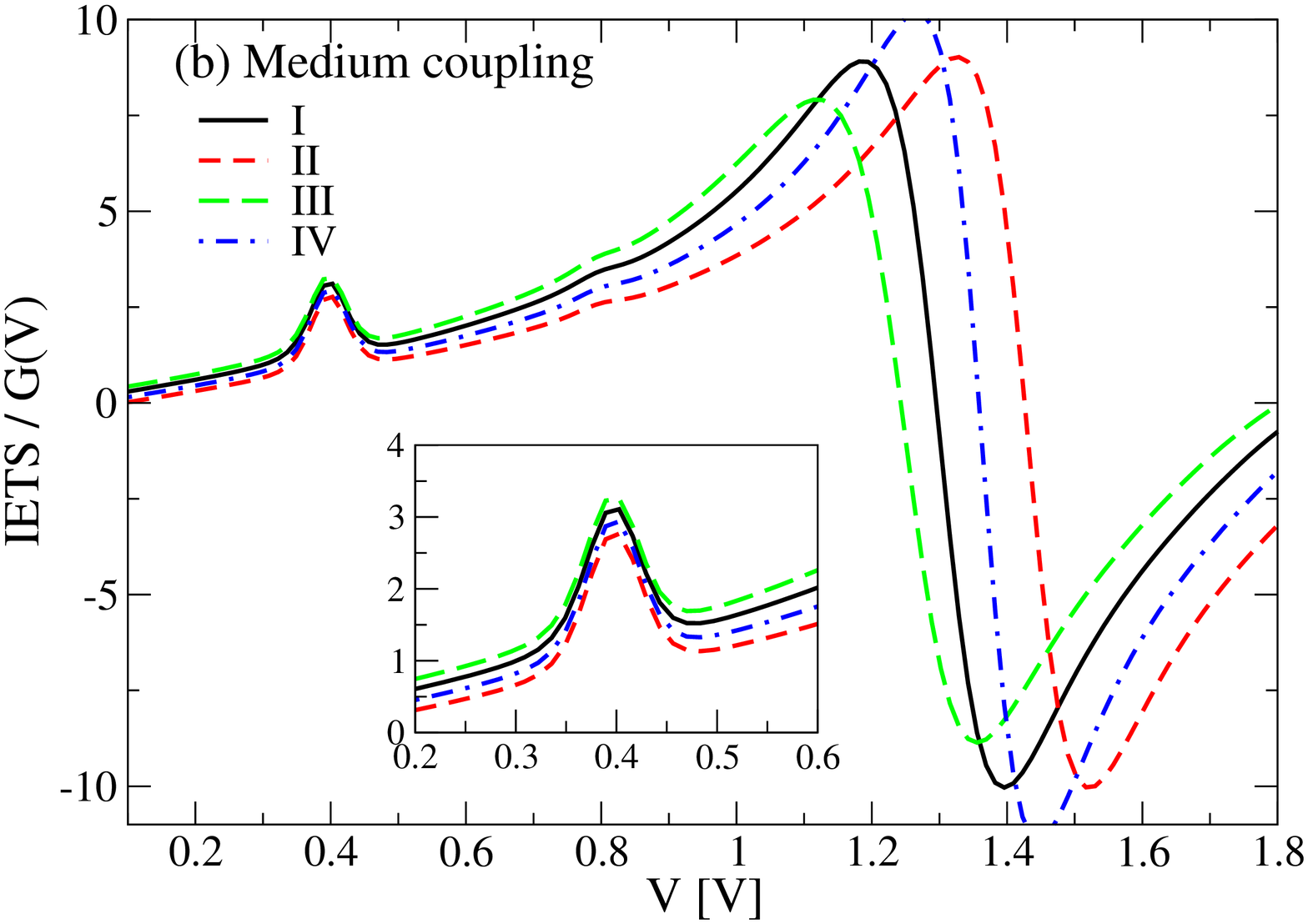} \\
\includegraphics[clip,width=0.8\columnwidth]{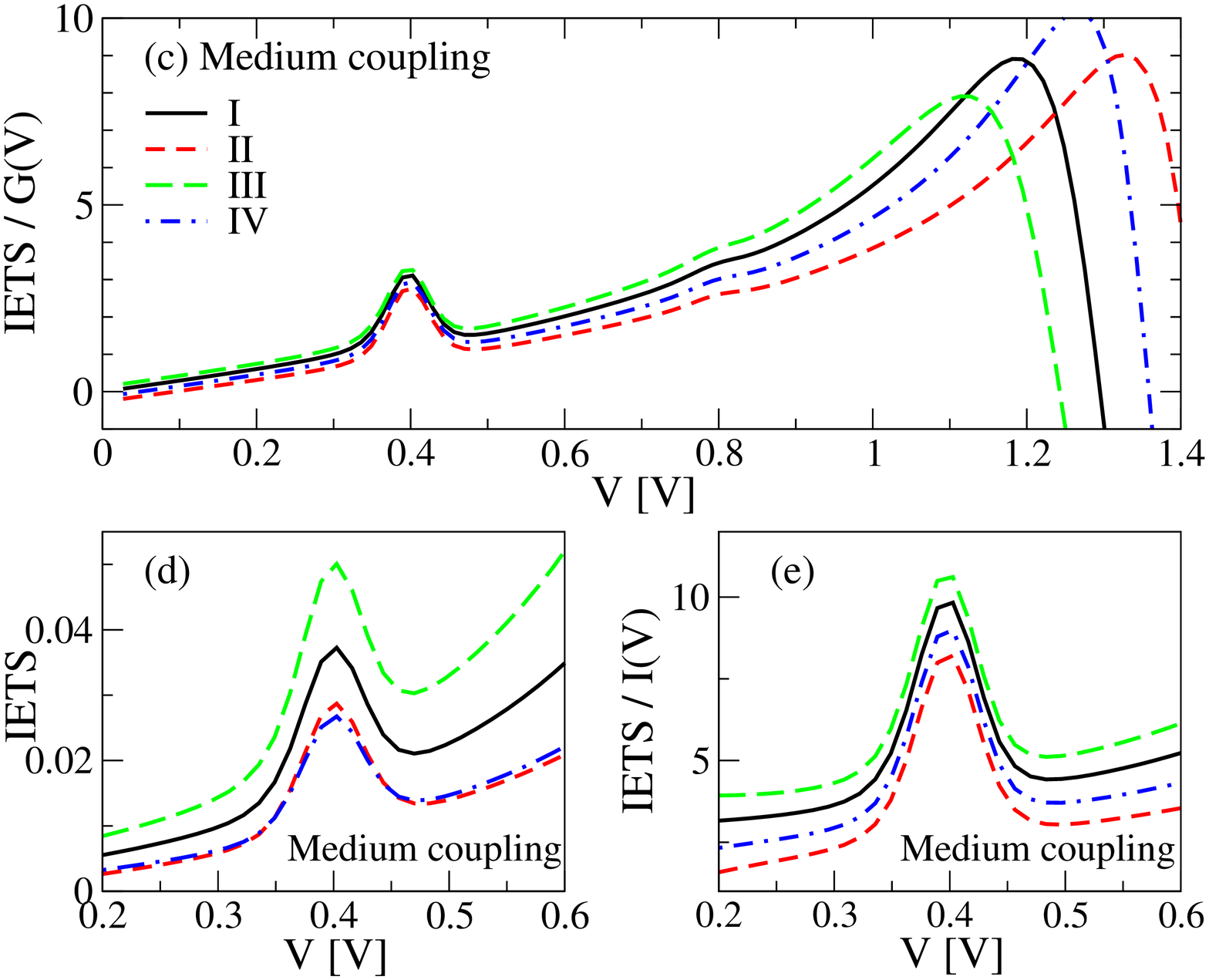} \\
\end{center}
\caption{IETS signal $d^2I/dV^2$ normalized by the conductance G(V)
  for the four cases depicted in fig.~\ref{fig:OurSystem}.  The top
  panel shows data for weak coupling to the leads ($t^S_{0L}=0.17$),
  the middle and bottom panels for medium coupling
  ($t^S_{0L}=0.17$). Both sets of calculations show that the position
  of the IETS feature at $V=\omega_0=0.4$ is not dependent on the
  fraction of potential drop at the contacts.  The bottom panels show
  different normalizations of the IETS as well as the bare IETS
  signal, as different normalization conventions are often used in the
  experiments.  There is no need to show $(d^2I/dV^2)/(I/V)$ since in
  this bias regime the current is linear $G\sim I/V$.}
\label{fig:IETS-curves}
\end{figure}

Results for the IETS signal for the same parameters are shown in
Fig.~\ref{fig:IETS-curves}, where we clearly observe a rather
different behavior in the conductance peaks.  Here we see two separate
features---the first is a peak at the vibration energy $\omega_0$, as
we would expect for the off-resonant regime
\cite{Galperin:2004b,Ness2010,Dash2011}. The amplitude of this
feature is proportional to $\gamma_0^2$ while the width is dependent
on the other parameters of the junctions
\cite{Paulsson:2008,Troisi:2005}. At higher biases we observe the
conductance peak, which occurs at the polaron-shifted electronic level
$\tilde\varepsilon_0$. Increasing the coupling to the leads yields an
effective decrease of the conductance peaks accompanied by an increase
in the width, as expected. This leads to better contrast in IETS
between the pure inelastic feature at $\omega_0$ and features
associated with elastic and inelastic resonant tunneling: the
amplitude of the inelastic feature is much less dependent on the
coupling to the leads.

It is also clear from Fig.~\ref{fig:IETS-curves} that the position
in bias of the IETS features around $V=\omega_0$ does not depend at
all on the strength of the coupling to the leads and on the fraction
of potential drop at the contacts.  In other words, the position of
the IETS feature is independent of the nature of the contacts. This is
the important result of this paper, as it means that the IETS signal
is more stable, in terms of spectroscopic information, than the
conductance itself.  The IETS signal from the internal vibrational
modes of the molecule is not strongly dependent on the way the
molecule is connected to the leads.

This occurs because, in spectroscopic terms, the IETS signal depends
only on the difference between the left and right Fermi levels. Thus a
vibration mode can only be excited by inelastic collision with the
charge carriers once the bias exceeds the vibrational energy, i.e.\ $V
\gtrsim \omega_0$, which leads to a corresponding feature in the IETS.
Note that this is not dependent on the model of the electron-vibron
coupling---it is purely an effect of non-equilibrium inelastic
transport at the threshold bias $V \gtrsim \omega_{\rm ext}$, where
$\omega_{\rm ext}$ could be any excitation of the system coupled to
the injected charge carriers.  There are also extreme asymmetric cases
for which the lineshape of IETS may vary or become a dip rather than a
peak \cite{Galperin:2004b}, but the position of the feature will
nonetheless remain at $V \sim \omega_0$.

The conductance peaks at $V \sim \tilde\varepsilon_0$, however,
correspond to resonant transmission through electronic levels, with or
without vibration replica\cite{Dash2010b}. Such resonances in the
conductance depend on the relative position of the two Fermi levels
with respect to the renormalized electronic levels, i.e. similar to
band offsets in semiconductors.  The conductance peak position is
therefore dependent on the fraction of potential drop at the
contacts. In our model, the left Fermi level moves up, while the right
Fermi level moves down, for positive bias.  The fraction of how much
$\mu_L$ moves up for a given bias is determined by the factor
$\eta_V$. The smaller $\eta_V$, the larger the bias $V$ has to be for
$\mu_L$ to become resonant with an electronic level
$\tilde\varepsilon_0$.  Such a mechanism explains the variation of the
conductance peak positions for the different cases (I-IV) of
fluctuations in the junction that we have considered in
Fig.~\ref{fig:OurSystem}.  These results confirm experimental common
knowledge: although an individual molecular device based on the use of
the conductance as the key signal may work well \cite{Quek:2009},
unless the fabrication of the device can be reliably controlled and
reproduced then mass production will remain impossible.  This is
especially true because the exact nature of the lead-molecule contacts
is unknown and as yet impossible to reproduce to specification.

The results we have to chosen to present in this section are obtained
for a variation of $10\%$ of the hopping integrals.  Obviously, larger
variations will lead to stronger effects, i.e.\ more important
modification of intensities of both the conductance peaks and the IETS
features, and more important shifts of the position of the conductance
peak. The IETS features will remain fixed at the same bias.  It
appears here that even small variations of the hopping integrals lead
to substantial effects.  As an illustration, let us recall that in
reality the hopping integrals between two different electronic
orbitals vary exponentially with the distance between the two atoms
supporting these orbitals.  The zero-point motion associated with
quantum fluctuations are of the order of 0.01-0.05 \AA, leading to
variations in the hopping integrals of a few percent or less. While
coordination-induced variations of order of 0.1 \AA~ in the
inter-atomic distance (roughly 10\% for a carbon double bond, well
beyond thermal fluctuations at room temperature) would lead to changes
of 20 to 50\% (and more) of the hopping integrals. Our choice of
$10\%$ variation of the hopping integrals is intermediate between
these two regime.

Furthermore, the analysis performed in this section is well suited
for large molecular system which have internal vibration
modes of atoms not involved in the bonding of the molecule to the
leads.
Indeed, for small molecules, most, if not all, of the atoms constituting 
the molecule are in close contact with the surface of one or both electrodes. 
In these cases the molecular vibronic properties (frequency of vibration and 
electron-vibration coupling matrix elements) are strongly dependent on the 
coupling of the molecule to the lead. All (or most) of the atoms involved 
in the chemical bonds inside the molecule are also involved in the bonding 
to the leads. These effects have been clearly shown in experimental
as well as theoretical works \cite{Hahn:2001,Tal:2008,Vitale:2010,
Garcia-Lekue:2011}.

However for larger molecules (molecular wires) it is clear that many vibration 
modes - the ones mostly located inside the molecular wire - will be much less 
dependent on the chemistry of bonding of the end atoms of the molecule to the 
electrodes. These modes are usually the optical-type modes and are strongly coupled 
to the LUMO and/or HOMO frontier orbitals as we show in the next section.

It is with these systems in mind that we aim to build functionality.
Hence the IETS signal of such systenes rids us of the need to control accurately 
the nature
of the contacts, and is thus a much more useful signal to consider
when designing and building functional single-molecule devices.  In
order to exploit this phenomenon and build a useful device, we need
some form of external control over the position and/or amplitude of
the IETS feature.  This external control could take several forms
(magnetic field, chemical concentration, pressure...), but here we
propose the use of an external electric field, in a similar way as a
gate voltage is used in Ref.~[\onlinecite{Song:2009}] to control the
conductance.

\section{Modifying the electronic and vibronic properties of a
  molecule with a gate potential}
\label{sec:AbInitio}

We now explore how the properties of the IETS signal we have
demonstrated in the previous section can be used for a more realistic
system.

For this we use \emph{ab-initio} calculations to study the effect of
an uniform electric field, acting as a gate voltage, on the electronic
and vibronic properties of a realistic molecular system.  We calculate
how the corresponding values of $\omega_0$ and $\gamma_0$ in our NEGF
calculations are modified by the external field for selected
vibrational modes coupled to the molecular \textsc{homo} and
\textsc{lumo} levels. We then demonstrate in the next section that
these results can be used to design a selectively functionalized
single-molecule device.

As full \emph{ab-initio} calculations for a realistic single-molecule
device with three terminals are extremely computationally demanding,
especially for self-consistent calculations, and are beyond present
computational power if one has to take the full non-equilibrium and
many-body effects into account, one has to introduce some
approximations.

In the following, as a first step of calculations to analyze the
potential of functional devices using the IETS signal, we perform the
calculation for an isolated molecule in the presence of an electric
field perpendicular to the backbone of the molecule.
However in real devices with metallic electrodes, the electrostatic 
potential acting on the molecule from an applied field perpendicular 
to the interelectrode spacing will be substantially distorted, on the
scale of the molecule, from an uniform field we use in the calculations. 
Hence the results we show in this section provide only the general trends
of the field dependence on the IETS signal for our model
\emph{ab-initio} calculations.

Furthermore, it is known that in real systems the presence of the electrode 
and the
coupling of the molecule to the electrodes are crucial to determine
accurately the charge transfer and transport properties of such a
molecular nanojunctions.  We have shown in the previous section that
the nature of the contacts (especially the strength of the couplings
and the corresponding potential drops) do indeed dominate the
properties of the conductance. The contacts also play an important 
role in other physical transport properties such as heat transport
\cite{Cuansing:2010}.

However, we have also shown that the IETS signal is virtually
independent of these characteristics of the contacts (hence the study
of an isolated molecule).  Furthermore, we concentrate our
calculations on the effects of the electric field on the vibron modes
which have a weak amplitudes at the ends of the molecules.  Normally,
such modes would not be strongly coupled to the electrodes if the
molecule were to be fully connected in a realistic molecular device.

\begin{figure}
\begin{center}
\includegraphics[clip,width=0.5\columnwidth]{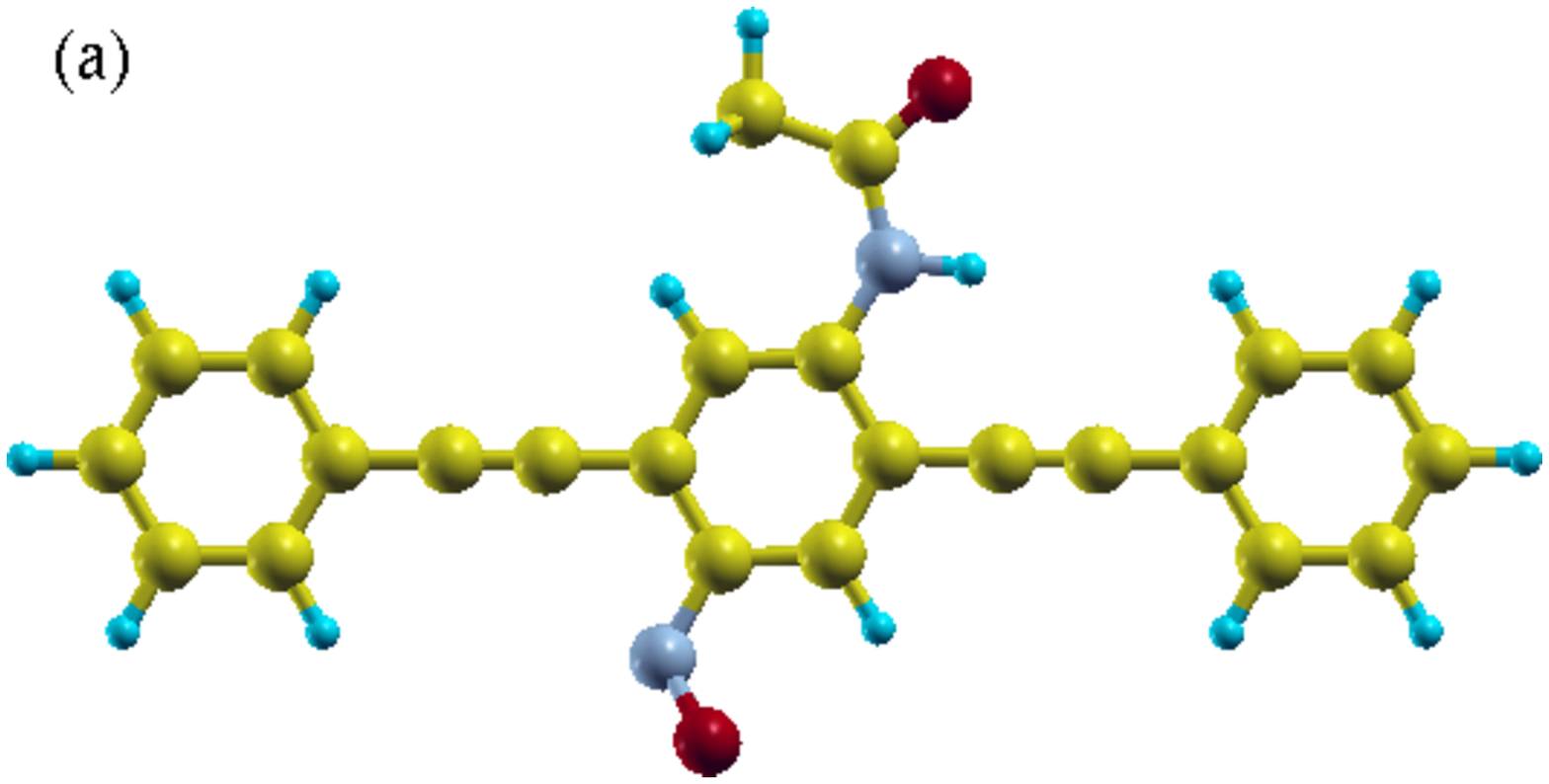}

\includegraphics[clip,width=0.5\columnwidth]{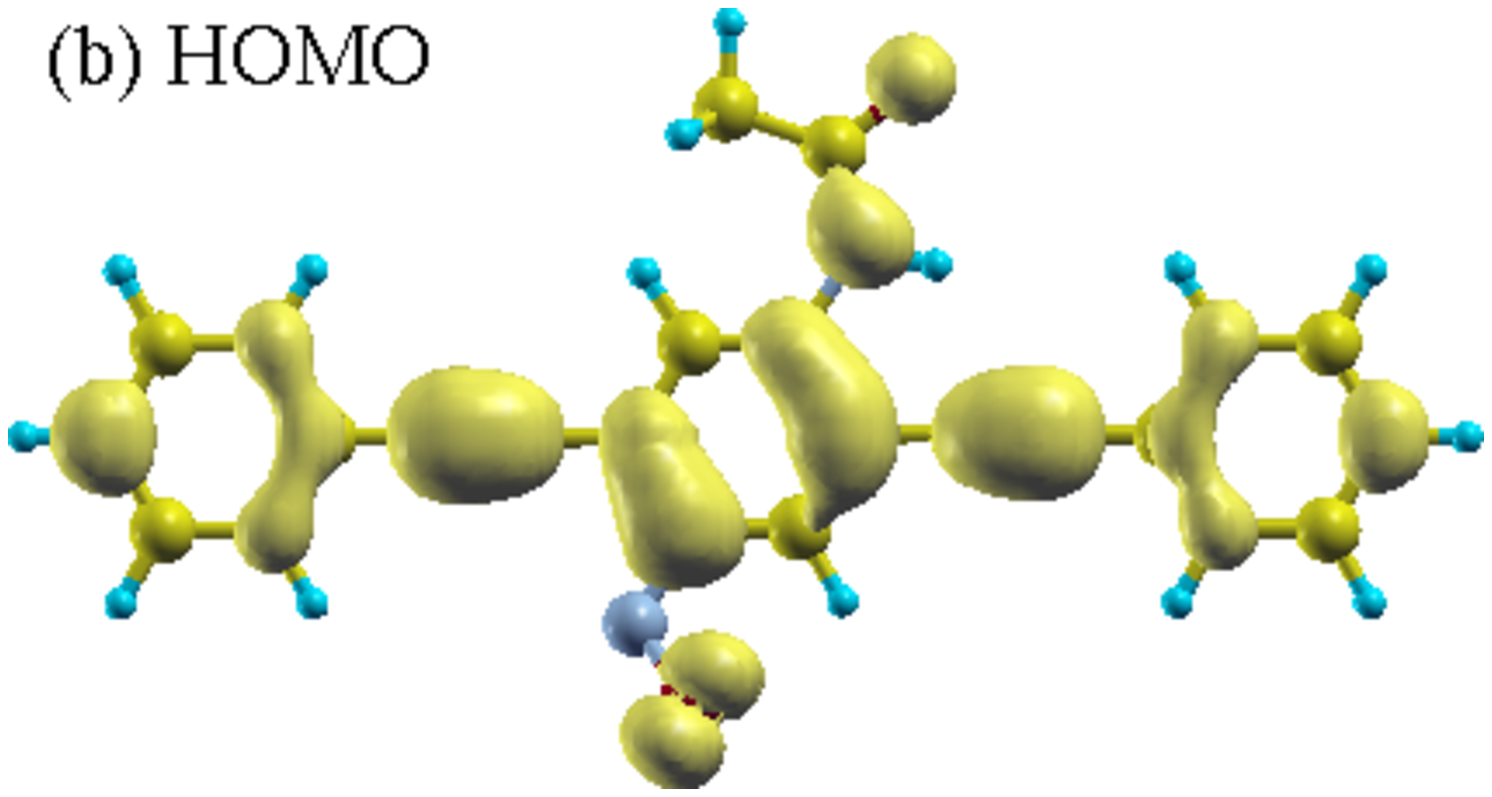}\hspace{8mm}\includegraphics[clip,width=0.5\columnwidth]{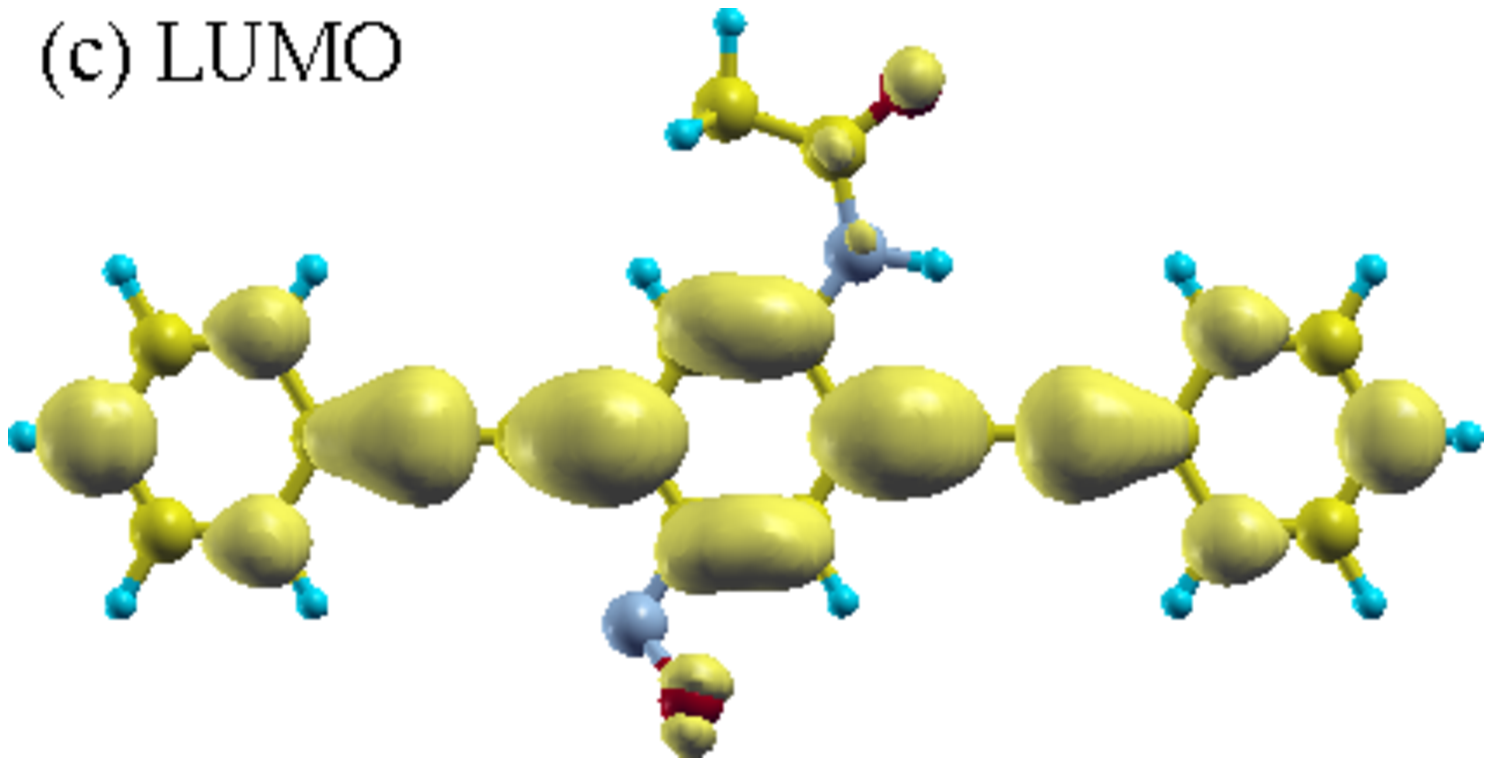}

\vspace{3mm}
\includegraphics[clip,width=0.5\columnwidth]{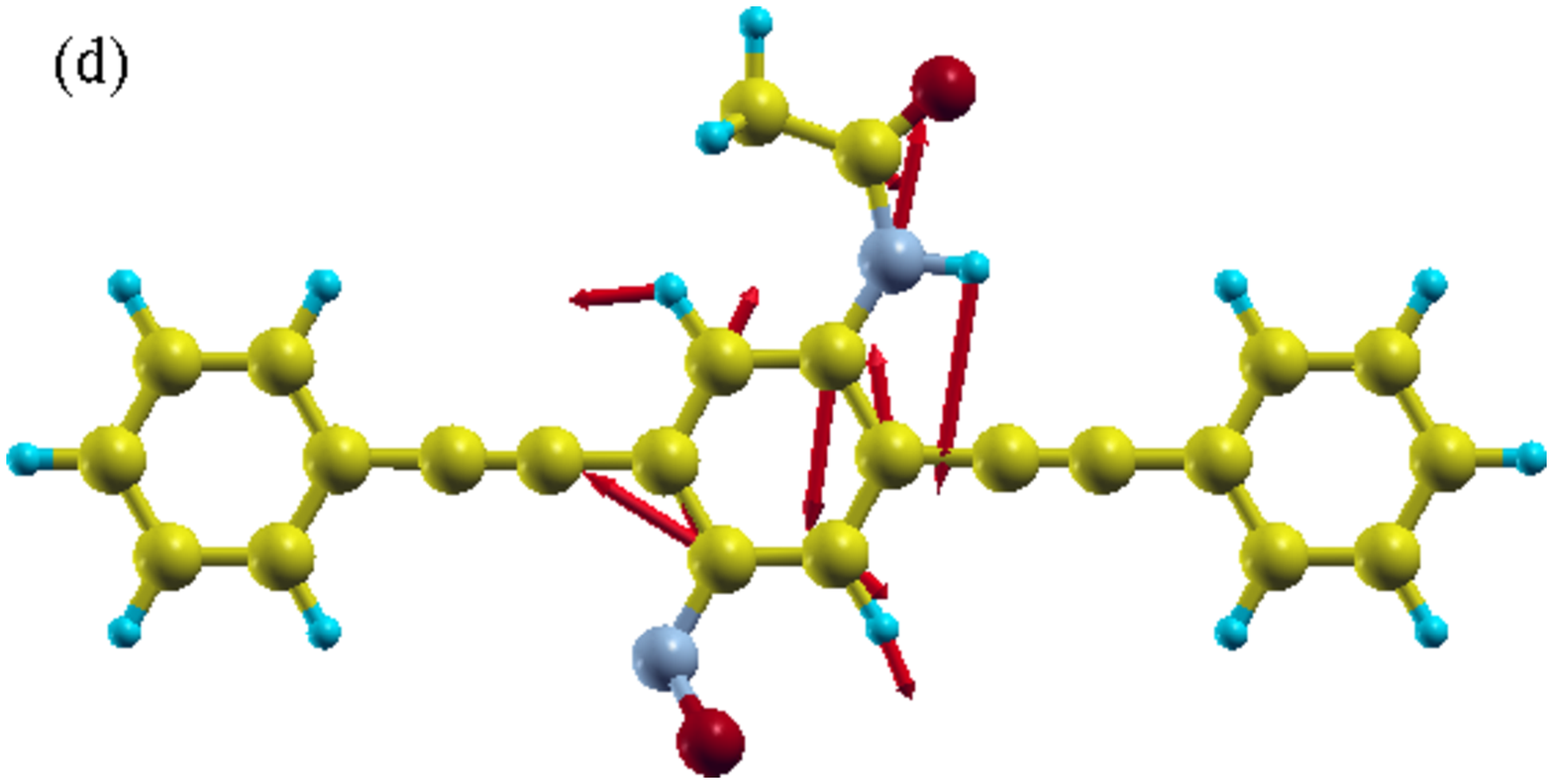}

\vspace{3mm}
\includegraphics[clip,width=0.5\columnwidth]{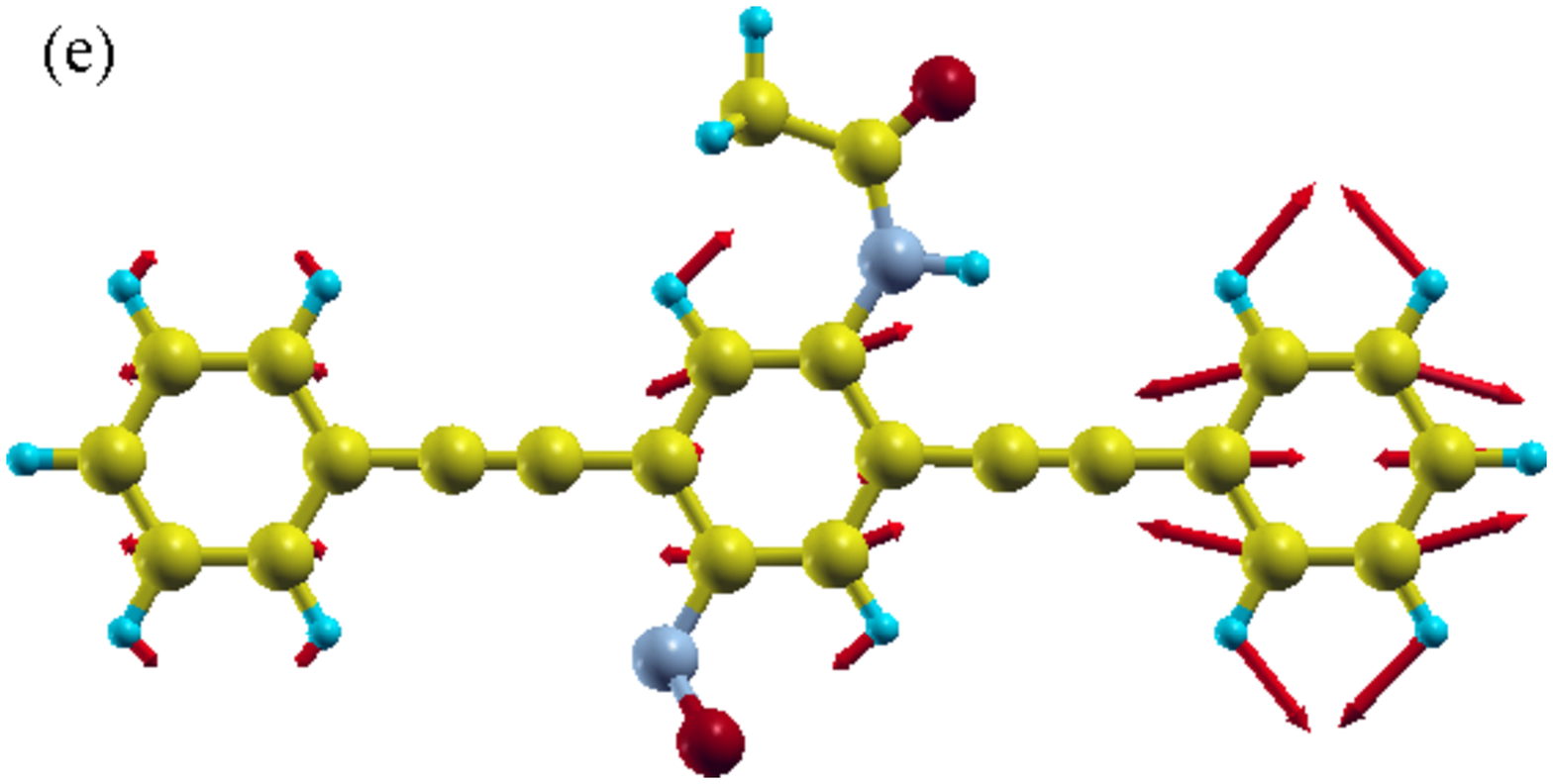}\hspace{5mm}
\includegraphics[clip,width=0.5\columnwidth]{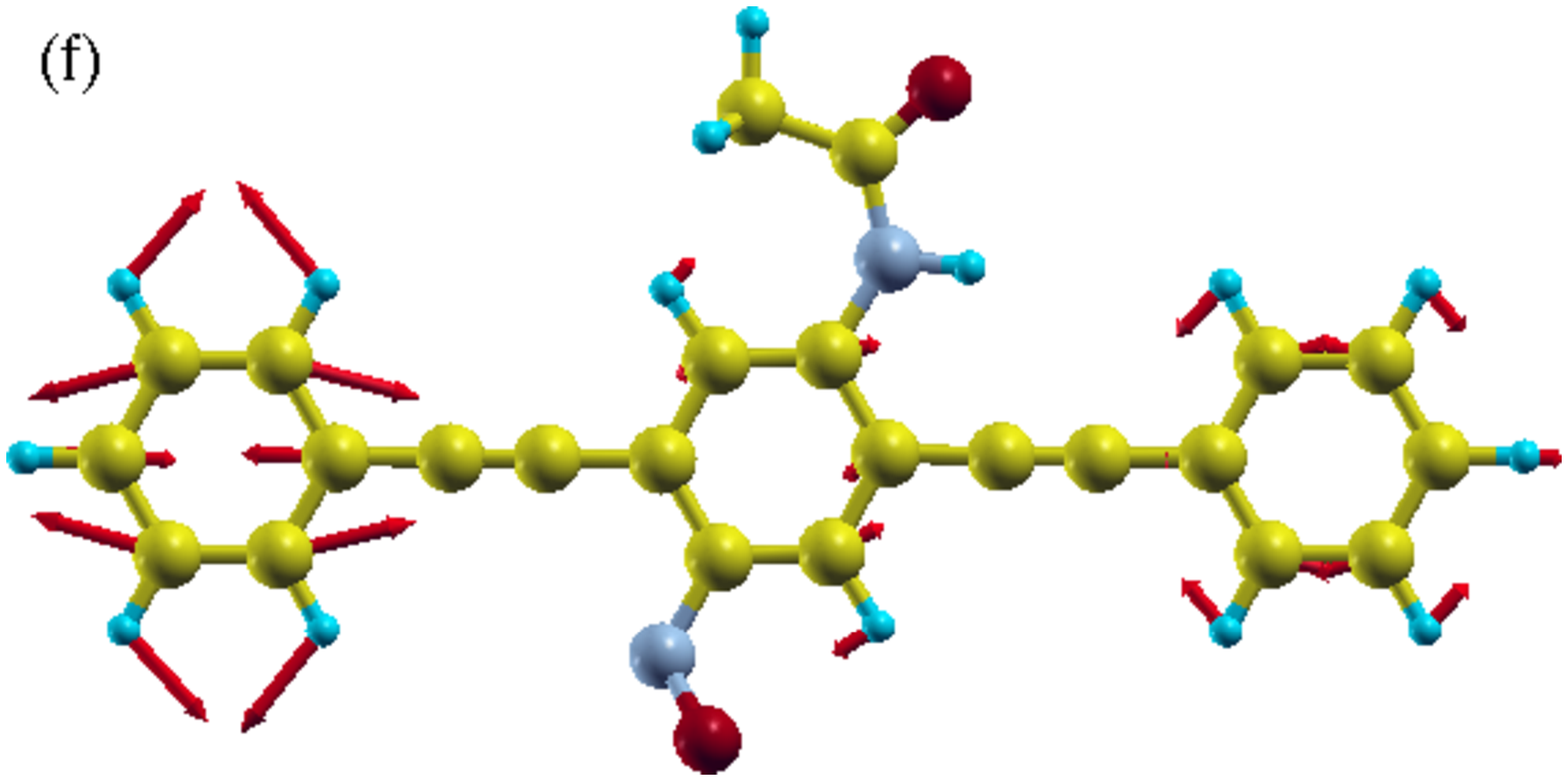}

\vspace{3mm}
\includegraphics[clip,width=0.5\columnwidth]{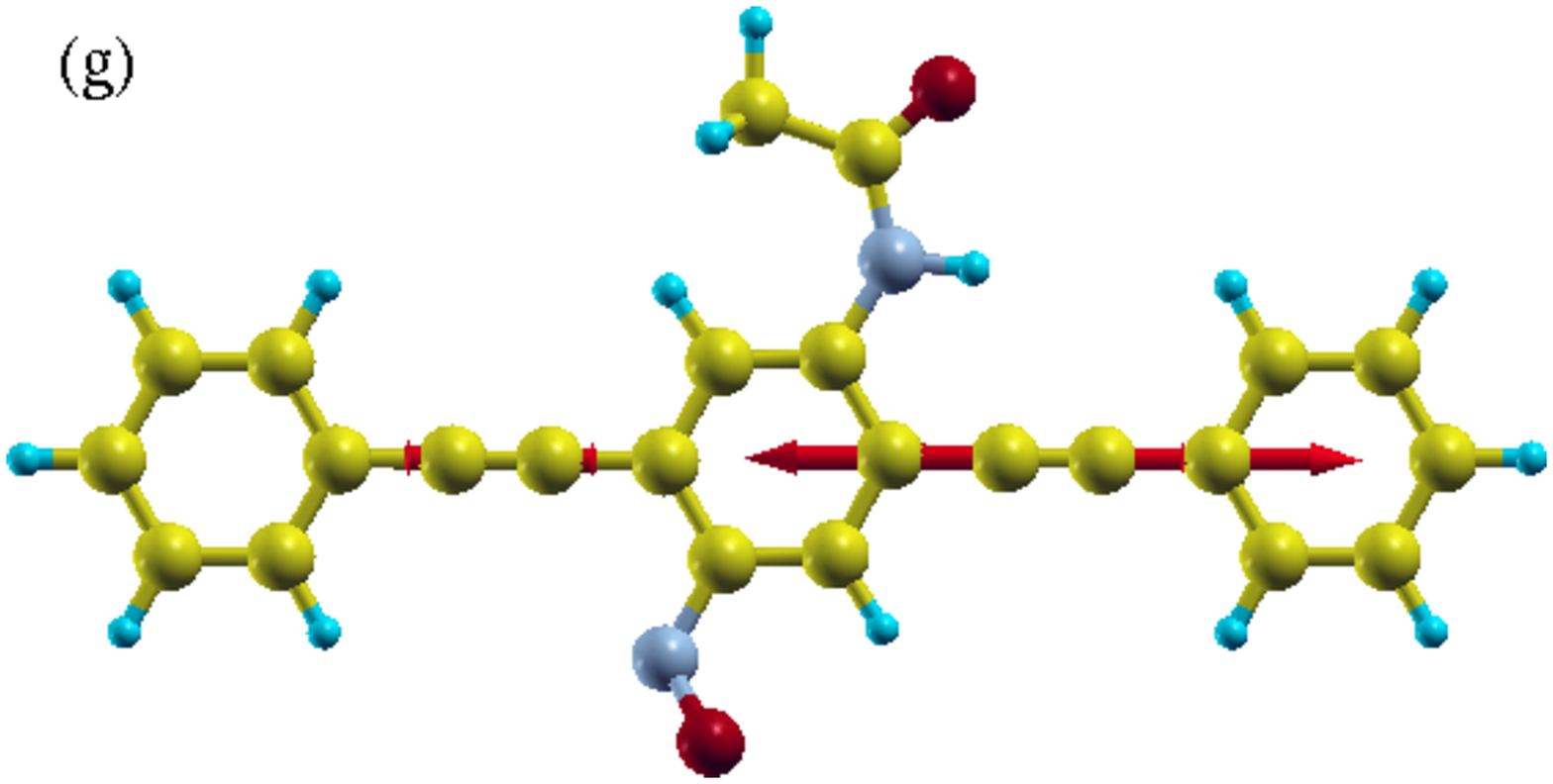}\hspace{5mm}
\includegraphics[clip,width=0.5\columnwidth]{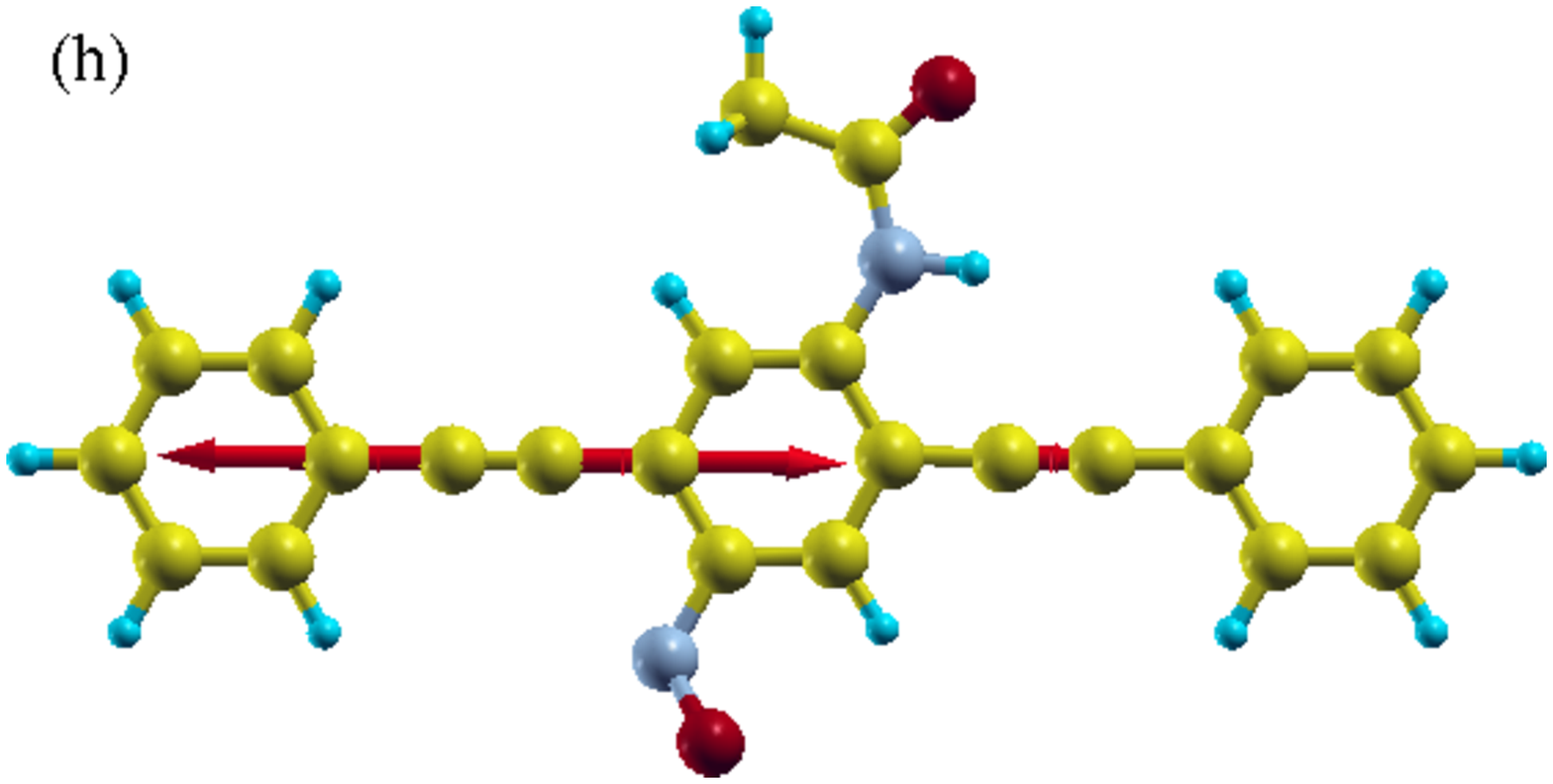}
\end{center}
\caption{Molecular wire: (a) ball and stick representation. (b) {\sc
    homo} state. (c) {\sc lumo} state. (d---h) Eigenmodes of
  vibration. (d) Mode strongly coupled to the {\sc homo}, mode
  $\lambda=111$ with $\omega_\lambda \sim 197$ meV; (e-f) Modes
  strongly coupled to the {\sc lumo}, mode (e) $\lambda=114$ with
  $\omega_\lambda \sim 206$ meV, and mode (f) $\lambda=115$ with
  $\omega_\lambda \sim 207$ meV; (g-h) Modes strongly coupled to both
  the {\sc homo} and {\sc lumo}, mode (g) $\lambda=118$ with
  $\omega_\lambda \sim 281$ meV, and mode (h) $\lambda=119$ with
  $\omega_\lambda \sim 284$ meV.}
\label{fig:molecule}
\end{figure}

As our test system, we choose the molecule
2,5-di[2'-(para-acetylmercaptophenyl)ethinyl]-4-nitro-acetylanilin,
shown in Fig.~\ref{fig:molecule}(a), which has previously been used
for transport measurements in a break junction \cite{Reichert:2003}.
Molecules with a similar ethynylphenyl-based backbone but with
different redox centers in the middle benzene ring have also been used
in self-assembled monolayer transport measurements
\cite{Chen_J:1999,Chen_J:2000}.  Another reason to chose this molecule
is based on the fact that the IETS signal of conjugated molecules can
be modified by a gate voltage has been shown in
Ref.~[\onlinecite{Song:2009}]. Furthermore our candidate molecule has
peripheral chemical groups that provide additional properties.

This molecule has specific properties that we are able to exploit.
Firstly, the form of the side-chains provide a permanent electric
dipole, so that the component of the electric field (associated with
the gate voltage) that is perpendicular to the backbone of the
molecule will show substantial effects. In particular, the electric
field will polarize the electron cloud along the dipole. In the regime
of strong field, the applied electric field might even bend the
molecular backbone.  Secondly, we will show that the dominant phonon
modes coupling to the \textsc{homo/lumo} levels are situated on the
central part of the molecule, and are thus effectively separated from
the leads.

Because of this, and the properties of the IETS signal versus the
nature of the contacts, we do not include leads in our
\emph{ab-initio} calculations. Rather, we replace the end groups of
the molecule by hydrogen atoms, and then fix these terminal atoms
within our supercell when structural relaxations are performed.

We calculate the ground-state electronic and vibronic properties of
the system using the ABINIT package \cite{abinit:2009}.  The
calculations are performed with Trouiller-Martin pseudopotentials,
using the local-density approximation (LDA) and the
exchange-correlation functional from S. Goedecker, M. Teter, and
J. Huetter \cite{Goedecker:1996}.  We use a supercell of size $50
\times 30 \times 50$ [bohr$^3]$ with a plane-wave cutoff of 30 Ryd at
the $\Gamma$-point.  The geometries of the molecule in the absence and
in the presence of the electrical field are fully relaxed until the
maximum force on each atom is less than 0.04 eV/\AA.  These are
sufficient for our analysis of the frequency variation induced by the
external field.

We first calculate the molecular structure without any applied field.
The relaxed geometry is planar in the $(x,y)$ plane.  The molecule has
a DFT \textsc{homo-lumo} gap of 2.29 eV and presents a permanent
dipole moment $\vec{d}=d_0 (x,y,0)$, lying in the plane of the
aromatic cycles, of magnitude $d_0=1.878$ D and direction
$(x,y)=(\frac{1}{2},\frac{\sqrt{3}}{2})$ where the x-axis is long the
molecular backbone.

The vibronic properties of the molecule are calculated from the the
dynamical matrix calculated in linear-response DFT (as in
Ref.~\onlinecite{Verstraete2006}):
\begin{equation}
\label{eq:dynmat}
D^{ij}_{\alpha \beta} = \frac{1}{\sqrt{M_i
    M_j}} \frac{\partial^2 E_{\rm tot}}{\partial r_{i\alpha} \partial
  r_{j\beta}} ,
\end{equation}
where $r_{i\alpha}$ is the displacement of atom $i$ (of mass $M_i$)
$\alpha$ is the reduced direction and $E_{\rm tot}$ is the total
energy obtained from the DFT calculation.

The eigenvalues of the dynamical matrix $D^{ij}_{\alpha \beta}$ give
the square of the vibron frequencies $\omega_\lambda^2$ while the
eigenvectors are the eigenmodes of vibration $V^\lambda_{{i\alpha}}$.
The electron-vibration coupling matrix elements are calculated as
\begin{equation}
\label{eq:gLkk}
\gamma^\lambda_{kk'} = \sum_{i,\alpha} \langle \phi_k\vert 
\sqrt{\frac{\hbar^2}{2\omega_\lambda}}\ \frac{\partial H}{\partial r_{i\alpha}} 
\vert \phi_{k'} \rangle
V^\lambda_{{i\alpha}} ,
\end{equation}
where $H$ is Hamiltonian of the molecule with atomic positions $r_{i\alpha}$
and $\vert \phi_{k} \rangle$ are
the corresponding eigenstates.

\begin{figure}
\begin{center}
\includegraphics[clip,width=0.9\columnwidth]{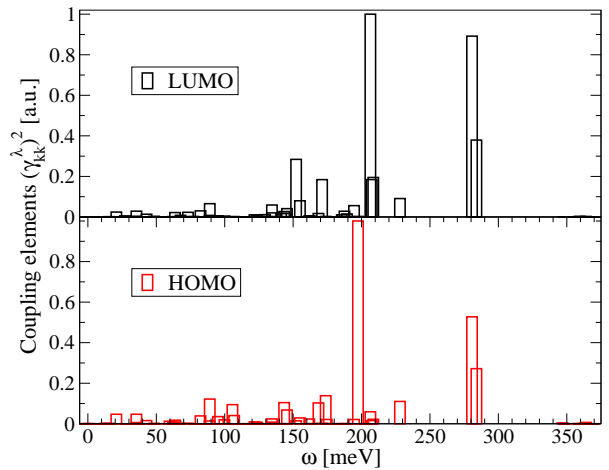}
\end{center}
\caption{Electron-vibration coupling: diagonal matrix elements $\gamma^\lambda_{kk}$ versus the vibration 
energy $\omega_\lambda$
for the two \textsc{homo} and \textsc{lumo} Kohn-Sham states $k$. The amplitude of the matrix elements 
are normalized to arbitrary units, i.e.\ ${\rm max}( \gamma^\lambda_{kk})^2 = 1$.
Note that since the IETS signal amplitude around $V=\omega_\lambda$ is proportional to $(\gamma^\lambda)^2$,
the graphs mimics the IETS signal in the low-bias region. 
One would get only two significant peaks in the range of applied bias $V=0$ to $V=0.35$ [V].}
\label{fig:HomoLumo-coupling}
\end{figure}

A ball and stick representation of the relaxed molecule, the
corresponding \textsc{homo} and \textsc{lumo} states, and the most
relevant vibrational modes are shown in Fig.~\ref{fig:molecule}.
The corresponding electron-vibration coupling matrix elements
$\gamma^\lambda_{kk}$ for the $k\equiv$\textsc{homo, lumo} states are
shown in Fig.~\ref{fig:HomoLumo-coupling}.  Clearly only a few of
the vibron modes couple strongly to the \textsc{homo} or \textsc{lumo}
states.

It should also be noted that, at the lowest-order in the
electron-vibron coupling, the amplitude of the IETS features is
proportional to $(\gamma^\lambda_{kk})^2$ as shown in Refs.~[
\onlinecite{Persson:1987,Paulsson:2005,Chen_Y:2005,
  Jiang:2005,Troisi:2005,Kula:2006}]. Therefore the graphs in
fig.~\ref{fig:HomoLumo-coupling} mimic the IETS spectrum in the range
of applied bias $V=0$ to $V=0.35$ [V].  In that range, one would only
see two main peaks around $V=0.20$ and $V=0.28$ [V].  Such a behavior
justifies {\it a posteriori} our NEGF analysis in terms of single-mode
excitation.

A central question is how the vibronic properties of the molecule
($\omega_\lambda$ for selected modes and $\gamma^\lambda_{kk}$ for the
same modes and for $k\equiv$ {\sc homo},{\sc lumo}) can be tuned.  In
our case the vibrations will be modified by applying an external field
$\vec{E}$ to the junction. The external field $\vec{E}$ acts as a
potential gate which may control the properties of the current flow
through the molecule \cite{Song:2009}.

Calculations of the electronic ground state and of the vibronic
properties of the molecule are performed in the presence of the
external electric field using the ABINIT package \cite{abinit:2009}.
Finite electric-field calculations, within periodic boundary
conditions, are performed by introducing an appropriate extra Berry
phase in the wavefunctions \cite{Nunes:1994,Nunes:2001,Souza:2002}.

In the present calculations, we take the electric field to be lying in
the plane of the aromatic cycles and to be perpendicular to the main
axis $x$ of the molecule, i.e.\ $\vec{E}=E_0 (0,1,0)$. In the full
three-terminal device, $\vec{E}$ represents the projection of the
three-dimensional electric field with the plane of the molecule and
perpendicular to its backbone. This is the component that is expected
to have the most efficient interaction with the dipole of the
molecule.  We chose that the magnitude of the field $E_0$ ranges from
0.001 to 0.020 a.u., i.e. from 0.051 up to 1.025 V/\AA.  This values
correspond to typical values of the electric field between a tip and a
sample within a scanning probe microscope setup \cite{Ness:1997b}.

\begin{figure}
\begin{center}
\includegraphics[clip,width=0.9\columnwidth]{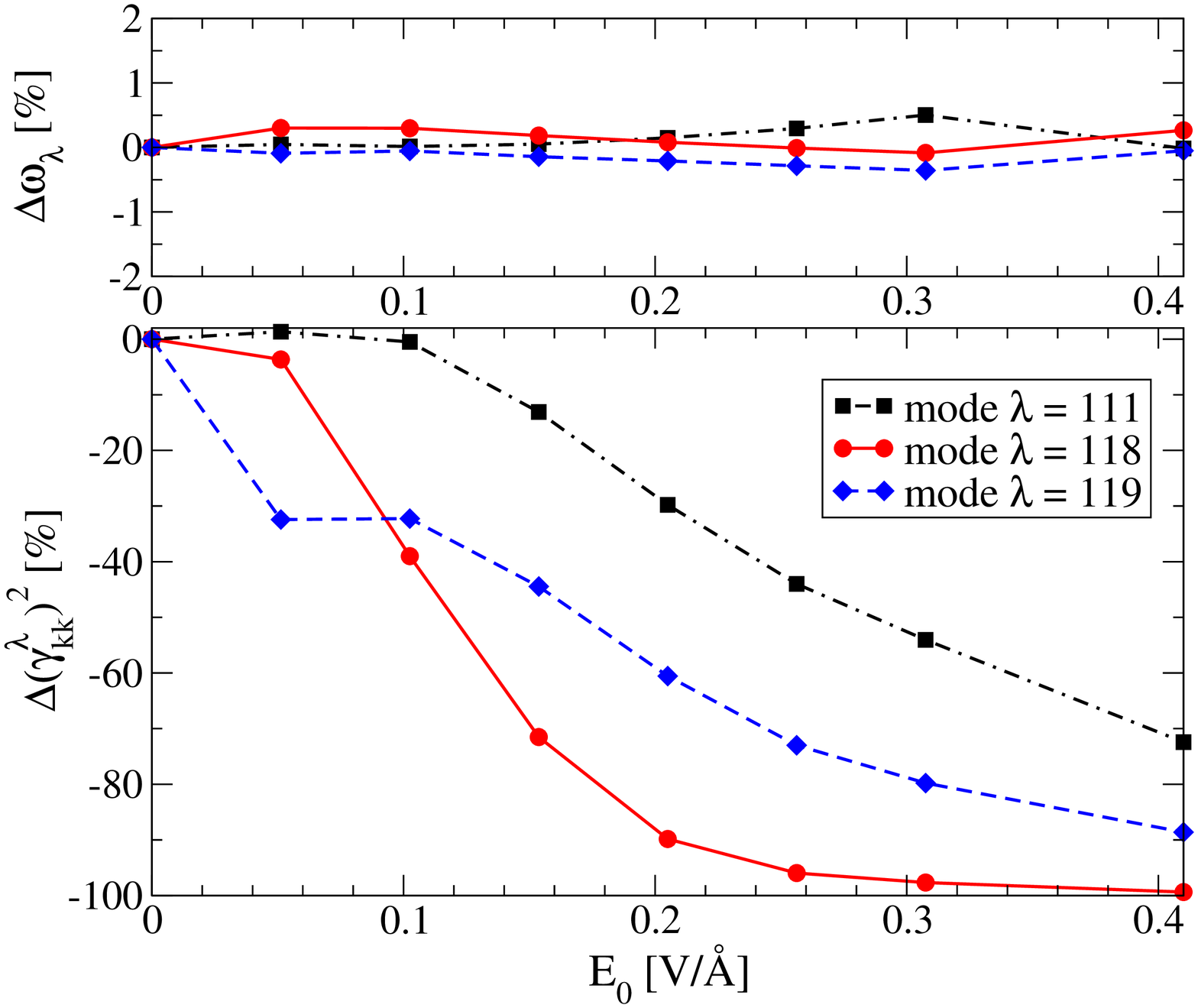}
\end{center}
 \caption{Relative evolution of the vibron energy $\omega_\lambda$  and
  coupling matrix elements $\gamma^\lambda_{kk}$ to the \textsc{homo} state
  as a function of the external electric field, 
  for modes $\lambda=111,\lambda=118$ and $\lambda=119$. 
  The zero-field values for the vibron energies and coupling constants (in meV) are 
  $\omega_\lambda$=197.39 and
  $\gamma^\lambda_{kk}$=54.14 ($\lambda=111$); $\omega_\lambda$=280.74 and
$\gamma^\lambda_{kk}$=39.32 ($\lambda=118$); $\omega_\lambda$=283.93 and
$\gamma^\lambda_{kk}$=28.23 ($\lambda=119$), respectively.
 }
  \label{fig:mode-vs-E_evolution}
\end{figure}

Fig.~\ref{fig:mode-vs-E_evolution} shows how the vibron energy and
matrix coupling element vary, for selected vibron modes, as the
external electric field is increased. We consider the vibron modes
which are the most strongly coupled to the frontier orbitals
(Fig.~\ref{fig:HomoLumo-coupling}).  We see that the coupling constant
$\gamma^\lambda_{kk}$ decreases monotonically for modes $\lambda$=111
and 119 over a range of $\Delta E_0=0.3$ [V/\AA] when $E_0\ge$ 0.1
[V/\AA], with an overall decrease of 70\% for mode 111, and 80\% for
mode 119.  Meanwhile mode 118 decreases much faster, showing a
switching-like behavior over a smaller range $\Delta E_0=0.15$ [V/\AA]
with an overall reduction of $\gamma^\lambda_{kk}$ larger than 90\%.
We also see that the applied external field lifts the degeneracy
between modes 118 and 119.

\begin{figure}
\begin{center}
\includegraphics[width=0.85\columnwidth]{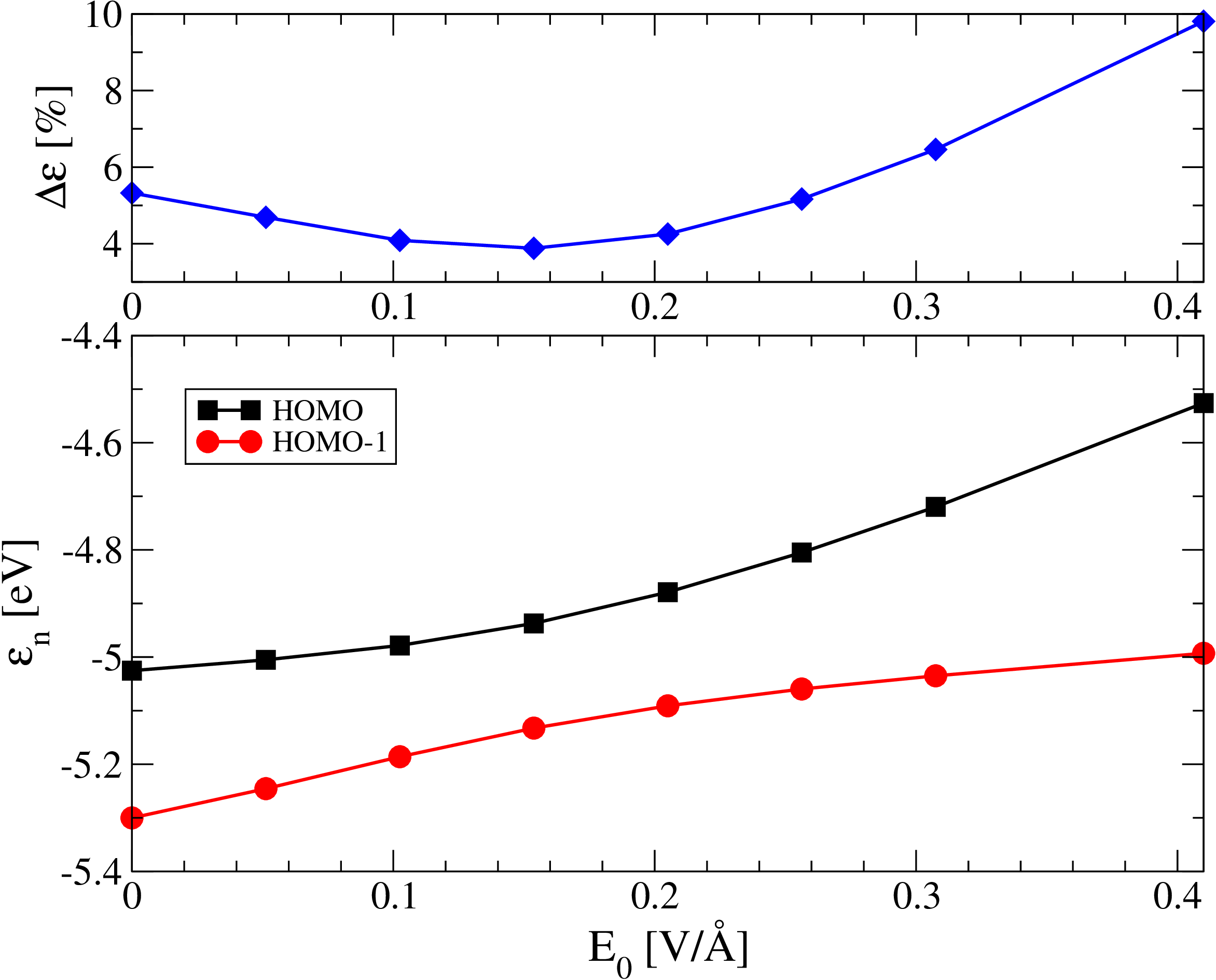}
\end{center}
\caption{Relative evolution of the {\sc homo} and  {\sc homo-1} electronic levels $\epsilon_n$ 
versus the applied external field. The zero field value of the {\sc homo} level is  
$\epsilon_{\rm HOMO}$=-0.18468 Ha = -5.0254 eV
and $\epsilon_{\rm HOMO-1}$=-0.19478 Ha = -5.3002 eV.
The top panel shows the relative energy separation between the  
{\sc homo} and  {\sc homo-1} levels 
$\Delta\varepsilon=(\varepsilon_{\sc homo}-\varepsilon_{\sc homo-1})/\bar\epsilon$ with
$\bar\varepsilon=\vert\varepsilon_{\sc homo}+\varepsilon_{\sc homo-1}\vert/2$.
Two linear regimes or a linear 
and weak quadratic regime are obtained for the dependence of $\epsilon_{\rm HOMO}$ on
the external field. The {\sc homo} level varies only by a few percent. The applied
electric field would not induce a transition from the off-resonant to the resonant
transport regime. Furthermore the energy separation with the {\sc homo} and  {\sc homo-1} 
levels increases with the applied field, validating even more the single-level
analysis used in our model calculations.}
\label{fig_EHOMO}
\end{figure}

In Fig.~\ref{fig_EHOMO}, we represent how the {\sc homo} and   
molecular level
varies with the applied electric field. We see that for the values of
the field we use, the variations of the {\sc homo} and {\sc homo-1} levels 
are only of a few percent. 
Assuming that in a three-terminal device, the Fermi level
at equilibrium is pined at the mid-gap of the molecule, the gate
voltage is modifying the position of the {\sc homo} level but does not
induce a transition between the off-resonant transport regime to the
resonant transport regime, i.e.\ the {\sc homo} level is always well
below the Fermi level.
Futhermore

\section{Application to functionality in single-molecule devices}

We now concentrate on the feasibility of obtaining a functional
single-molecule device by using the IETS signal.  The results of the
DFT \emph{ab-initio} calculations given in the previous section will
be used as input parameters in our NEGF model.  The vibron frequency
is $\omega_0 \leftarrow \omega_\lambda$, and the electron-vibron
coupling constant $\gamma_0^2 \leftarrow (\gamma^\lambda_{kk})^2$.  We
take the molecular electronic level $\varepsilon_0$ to be mid-gap,
i.e.~$\pm
(\varepsilon_{\textsc{LUMO}}-\varepsilon_{\textsc{HOMO}})/2$, which is
$\sim \pm 1.15$ eV.  Finally, we introduce an effective broadening
$t_{0L,R}$ of the molecular levels corresponding to the coupling with
the leads. The coupling is chosen such that the molecular level
broadening is less than the spacing between levels $(\varepsilon_{\rm
  HOMO}-\varepsilon_{\rm HOMO-1})/3.3 \sim$ 80 meV.

\begin{figure}
\begin{center}
\includegraphics[clip,width=0.9\columnwidth]{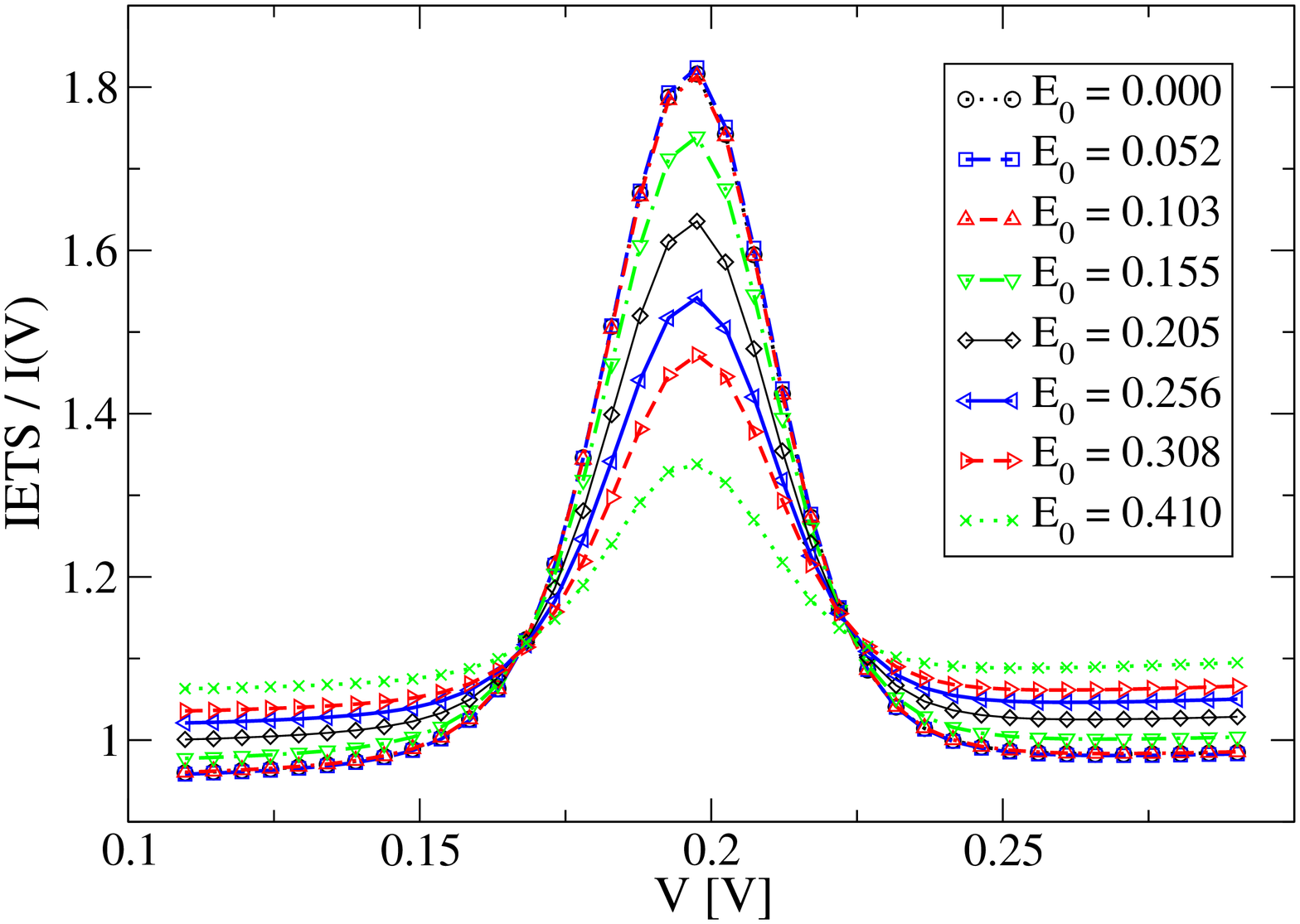}
\includegraphics[clip,width=0.9\columnwidth]{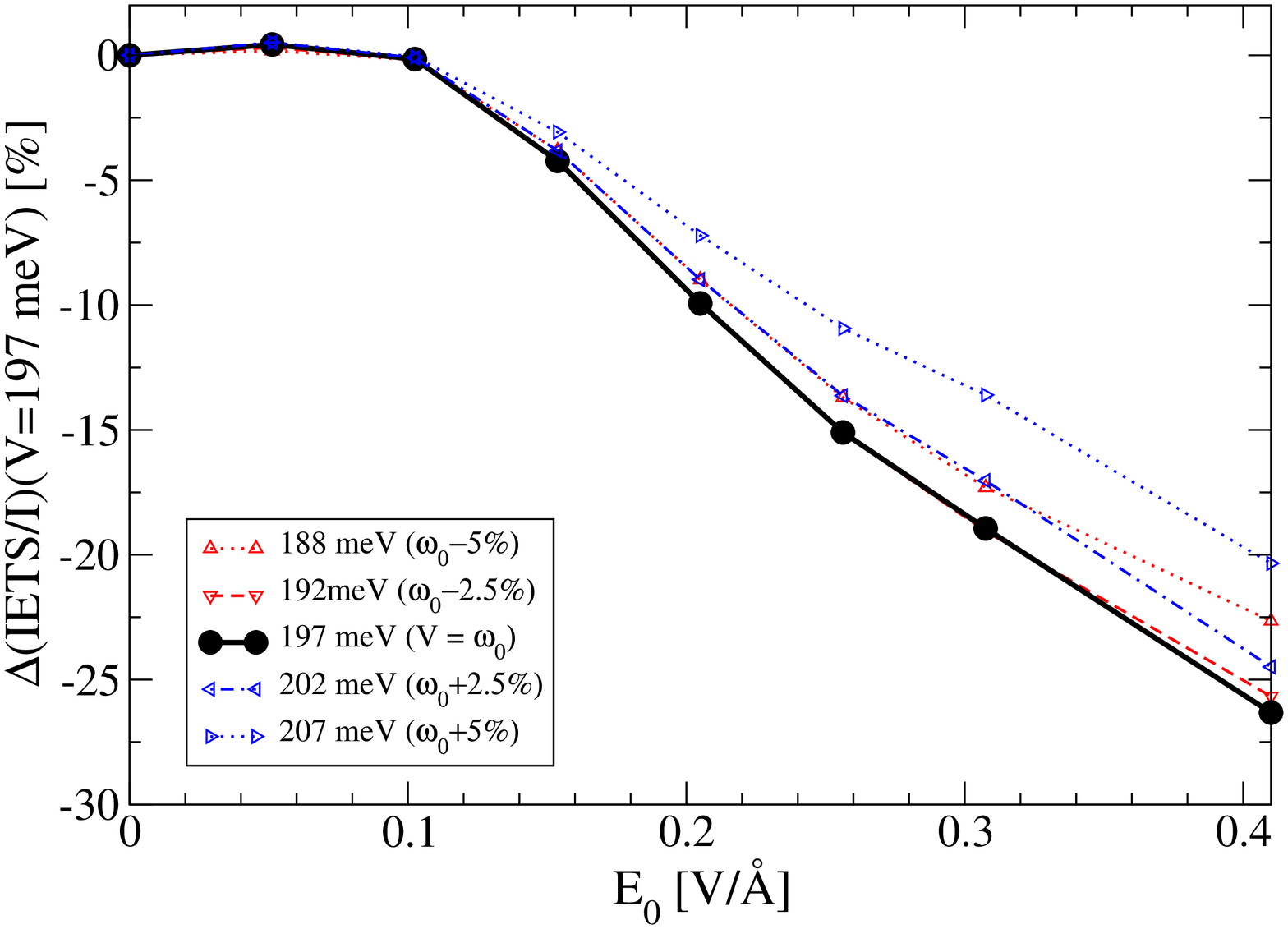}
\end{center}
\caption{Top panel: Evolution of the normalized IETS signal around
  $\omega_\lambda$=197 meV for mode $\lambda=111$ for the different
  values of external electric field $E_0$ [V/\AA]. The IETS signal is
  normalized by the value of the current $I(V)$ taken at the same
  bias. The inelastic peak is centered on $\omega_\lambda$=197 meV for
  all electric field values, only the peak amplitude and background
  changes.  Bottom panel: Relative evolution of the normalized IETS
  signal at the inelastic resonance $V=\omega_\lambda$=197 meV versus
  the values of external electric field $E_0$.  The evolution of the
  signal versus the gate voltage (related to $E_0$) is typical of a
  non-linear amplification process when $E_0$ decreases from a finite
  value to zero. }
\label{fig:IETSmode111_vsEfield}
\end{figure}

\begin{figure}
\begin{center}
\includegraphics[clip,width=0.9\columnwidth]{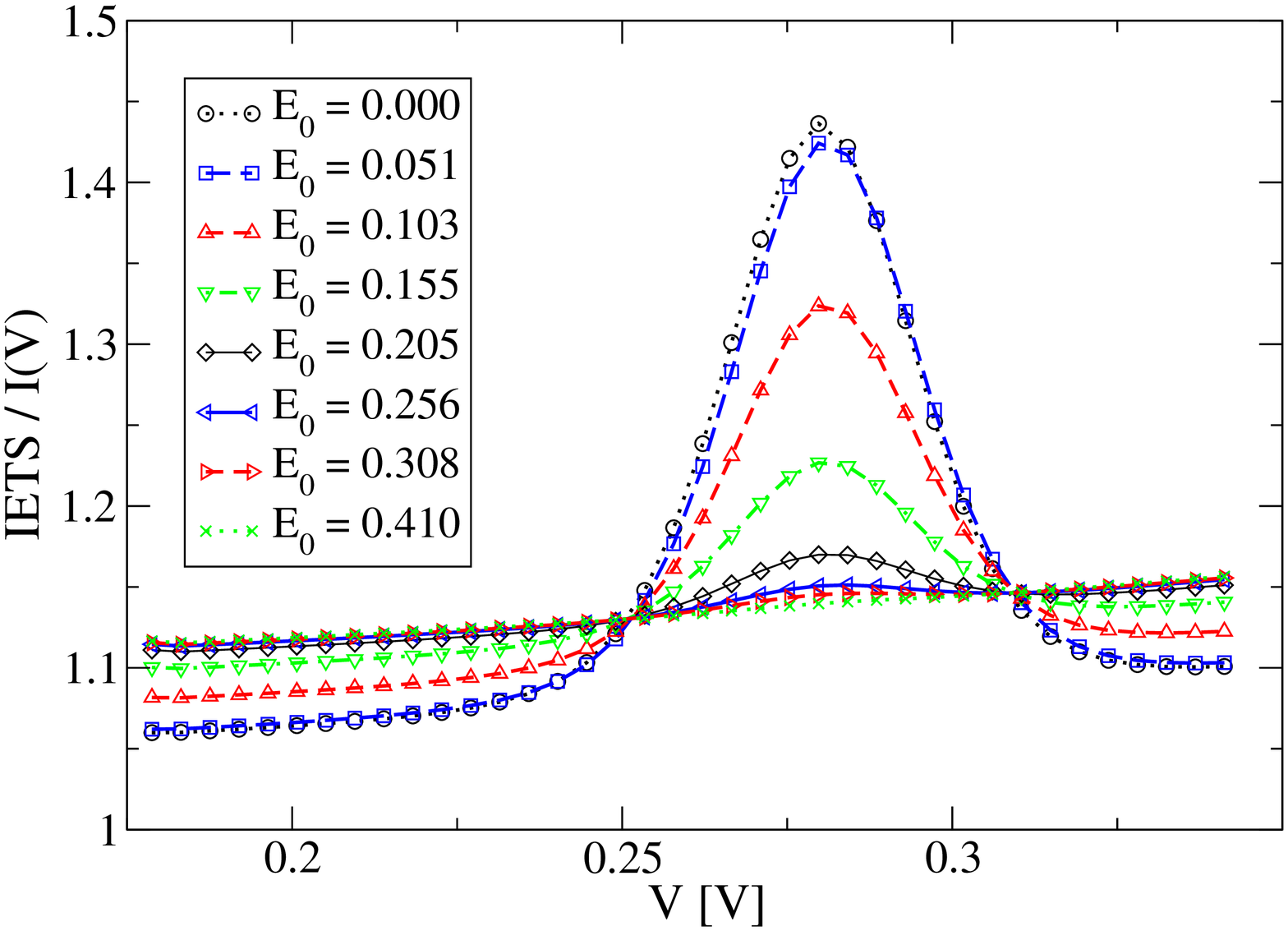}
\includegraphics[clip,width=0.9\columnwidth]{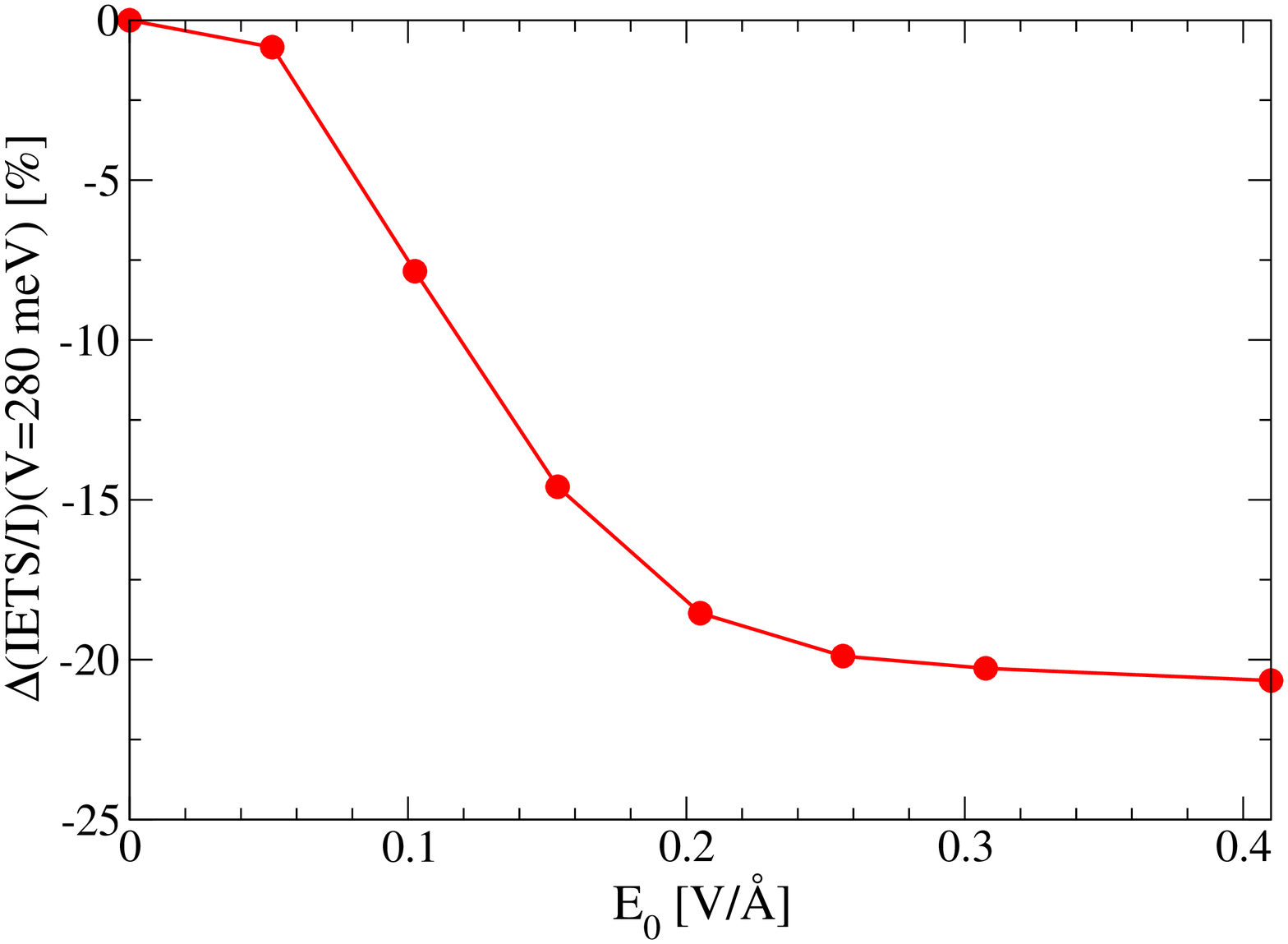}
\end{center}
\caption{
Top panel:
Evolution of the normalized IETS signal around $\omega_\lambda$=280 meV for mode $\lambda=118$ 
for the different values of external electric field $E_0$.
Bottom panel:
Relative evolution of the normalized IETS signal at $V=\omega_\lambda$=280 meV 
versus the values of external electric field $E_0$. 
The evolution of the signal versus the gate voltage (related to $E_0$) is
typical of an strong switching process from high to low values when
$E_0$ increases from 0.05 to 0.20 [V/\AA].
}
\label{fig:IETSmode118_vsEfield}
\end{figure}

Using these values, we calculate the IETS signal around $V \sim
\omega_0$, and study how this signal is modified by the external
applied field $\vec{E}=E_0 (0,1,0)$.  The upper panels of
Figs.~\ref{fig:IETSmode111_vsEfield} and
\ref{fig:IETSmode118_vsEfield} show the evolution of the the IETS
signal for NEGF calculations with the parameters for mode 111 and mode
118. Here we have chosen to normalize the IETS by the current itself,
as it reduces the effect of the slope background on the IETS peaks
(see Fig.~\ref{fig:IETS-curves}).

The lower panels of Figs.~\ref{fig:IETSmode111_vsEfield} and
\ref{fig:IETSmode118_vsEfield} show the relative evolution of the
normalized IETS signal for a fixed source-drain applied bias versus
the external electric field acting as a gate voltage.  These results
demonstrate that we can indeed achieve a functional behavior from the
IETS signal when working at low applied bias around the frequencies of
the vibron modes (here modes 111 and 118) that are strongly coupled to
the {\sc homo} molecular level.

The behavior of the IETS signal in Fig.~\ref{fig:IETSmode111_vsEfield}
is typical of a non-linear amplification of the source-drain current
versus the gate voltage represented by $E_0$. The signal is amplified
in a non-linear manner when $E_0$ decreases from a finite value to
zero. Note that the curve is simply convex with no inflection point.

In Fig.~\ref{fig:IETSmode118_vsEfield} we obtain a typical
switching behavior of the IETS signal from two plateaux of high and low
value as $E_0$ increases.  Note that now the curve presents an
inflection point around $E_0$=0.10 [V/\AA]; and that the switching
occurs over a small range of applied field $\Delta E_0$=0.15 [V/\AA].

The relative variations of the current-normalized IETS signal is of
the order of 20--25\% and is not as important as the relative variation
of the corresponding coupling matrix elements
$(\gamma^\lambda_{kk})^2$ shown in Fig.~\ref{fig:mode-vs-E_evolution}.
This is to be expected, since the full non-equilibrium inelastic
transport properties are not simply related in a linear way to the
square of the coupling matrix elements $(\gamma^\lambda_{kk})^2$, as
would be obtained from perturbation theory \cite{Chen:2004}.  In a
full non-equilibrium calculation, complex non-linear effects enter
into account \cite{Galperin:2004b,Egger:2008} and cannot be described
perturbatively.

In terms of practical devices, the IETS signal could be measured
by using an electronic circuit similar to that developed for non-linear
amplification through single fullerene molecules
\cite{Joachim:1997}.  However, in order to extract the
IETS, one would use a lock-in technique for the input source-drain
bias, and measure the second harmonic of output voltage at the load
resistance \cite{Hipps:1993}. One then gets a signal proportional to
the IETS signal since the device works within the linear regime for
the source-drain current versus source-drain bias for low applied bias
around the vibron frequencies (100-350 meV).

Until now, IETS measurements in molecular nanojunctions have been
performed at low temperature \cite{Liu:2004,Kushmerick:2004,
  Yu:2004,Wang:2004,Yu:2006,Chae:2006,Beebe_JM:2007,Okabayashi:2008,Kim:2010}.
This is necessary in order to have good mechanical stability of the
molecular junctions, as well as a good resolution of the IETS feature
which depends on the thermal fluctuations of the soft acoustic-like
vibration modes\cite{Ness:2006}. The width of the IETS features
increases with increasing temperature, with a loss of resolution above
liquid nitrogen temperature \cite{Wang:2004,Song:2009}.  All this
limits the use of the single-molecule devices to temperatures below
$\sim$70K.  It can be expected that by working with other candidates
of molecular connectors, or by using different concepts to build
three-terminal devices (i.e.\ for example by using carbon-based
electrodes \cite{Tuukkanen:2009}), this temperature limit could be
lifted.

\section{Discussion}
\label{sec:Conclusions}

Using a two-step theoretical framework we have shown that the IETS
signal through molecular junctions can be used to achieve
functionality within a single-molecule device.  By using a NEGF
approach for a model system, we have shown that the IETS is virtually
independent of the nature of the contact between the molecule and the
source and drain.  The electronic and vibronic properties of a
realistic molecular candidate (with an ethynylphenyl-based backbone)
are calculated using an {\em ab-initio} method.  We have shown how an
externally applied electric field, simulating the presence of a gate
electrode, can control the vibron properties of the molecule and
therefore the corresponding IETS signal.

Multi-functionality is demonstrated within a \emph{single} molecule:
non-linear amplification and switching are both present for different
vibron modes in the molecular system we have studied. Such
functionality should be reproducible from device to device, since the
IETS is virtually independent of variations in the molecule-lead
contacts.

Furthermore, a recent theoretical study on the difficulty of gate
control in molecular transistors \cite{Hou:2011} has shown that when
the molecular energy levels are away from the Fermi level (i.e. the
off-resonant transport regime), they can be shifted by the gate
voltage. However when the molecular levels move close to the Fermi
level, the shifts become extremely small and almost independent of the
gate voltage. This indicates that it may be difficult to use the gate
voltage to control transition in the conductance between the
off-resonant regime (``OFF'' state) and the resonant regime (``ON''
state).  This difficulty does not occur in our scheme, since we do
not rely on the use of the conductance as the functional signal.

In the present study we have considered the low electric-field regime
for the control of the IETS at finite bias.  In this regime, most of
the functionality comes from the dependence of the coupling matrix
elements $\gamma^\lambda_{kk}$ on the electric field, while the vibron
frequencies are more or less constant.  In the regime of stronger
electric field, the polarization of the electronic clouds may be strong enough
to lead to modification of some chemical bonds in the molecule. We
expect then that the variation of the vibron frequencies will be
important and the displacement of the vibron frequencies will dominate
the functionality of the single-molecule device.  

We can now suggest designing single-molecule sensors in the following
manner: one uses a molecule with peripheral chemical groups that can
actively react with surrounding molecules. After chemical reaction,
the vibron modes of the molecular backbone that couple strongly to the
{\sc homo}-{\sc lumo} molecular states, and/or the corresponding
coupling matrix elements, are modified. One can then monitor the state
of the molecular sensor by measuring the evolution of the IETS signal.
Finally, the mechanisms we describe above are most relevant in the
vibrational theory of the sense of smell (olfaction)
\cite{Turin:2002,Brookes:2007,Franco:2011}.

\begin{acknowledgments}

  This work was funded in part by the European Community's Seventh
  Framework Programme (FP7/2007-2013) under grant agreement no 211956
  (ETSF e-I3 grant). Part of the calculations were performed on
  Magerit (Red Espanola de Supercomputacion).

\end{acknowledgments}


\end{document}